\documentclass[twocolumn]{aastex63}
\usepackage{apjfonts}

\newcommand{\overbar}[1]{\mkern 3.0mu\overline{\mkern-3.0mu#1\mkern-3.0mu}\mkern 3.0mu}

\shorttitle{Host Dark Matter Halos of Red and Blue QSOs}
\shortauthors{Petter et al.}
\accepted{2022 January 18 to \apj}
\graphicspath{{./}{figures/}}
\begin{document}

%\title{Host Dark Matter Halos of SDSS Red and Blue Quasars}
\title{Host Dark Matter Halos of SDSS Red and Blue Quasars: No Significant Difference in Large-scale Environment}

\correspondingauthor{Grayson Petter}
\email{Grayson.C.Petter.GR@dartmouth.edu}

\author[0000-0001-6941-8411]{Grayson C. Petter}
\affil{Department of Physics and Astronomy, Dartmouth College, 6127 Wilder Laboratory, Hanover, NH 03755, USA}

\author[0000-0003-1468-9526]{Ryan C. Hickox}
\affiliation{Department of Physics and Astronomy, Dartmouth College, 6127 Wilder Laboratory, Hanover, NH 03755, USA}

\author[0000-0002-5896-6313]{David M. Alexander}
\affiliation{Centre for Extragalactic Astronomy, Department of Physics, Durham University, South Road, Durham, DH1 3LE, UK}

\author[0000-0003-4964-4635]{James E. Geach}
\affiliation{Centre for Astrophysics Research, School of Physics, Astronomy \& Mathematics, University of Hertfordshire, Hatfield, AL10 9AB, UK}

\author{Adam D. Myers}
\affiliation{Department of Physics and Astronomy, University of Wyoming, Laramie, WY 82071, USA}

\author[0000-0002-0001-3587]{David J. Rosario}
\affiliation{School of Mathematics, Statistics and Physics, Newcastle University, Newcastle upon Tyne, NE1 7RU, UK}
\affiliation{Centre for Extragalactic Astronomy, Department of Physics, Durham University, South Road, Durham, DH1 3LE, UK}

\author[0000-0003-1251-532X]{Victoria A. Fawcett}
\affiliation{Centre for Extragalactic Astronomy, Department of Physics, Durham University, South Road, Durham, DH1 3LE, UK}

\author[0000-0001-9307-9026]{Lizelke Klindt}
\affiliation{Centre for Extragalactic Astronomy, Department of Physics, Durham University, South Road, Durham, DH1 3LE, UK}

\author[0000-0002-8571-9801]{Kelly E. Whalen}
\affil{Department of Physics and Astronomy, Dartmouth College, 6127 Wilder Laboratory, Hanover, NH 03755, USA}



\begin{abstract}

The observed optical colors of quasars are generally interpreted in one of two frameworks: unified models which attribute color to random orientation of the accretion disk along the line-of-sight, and evolutionary models which invoke connections between quasar systems and their environments. We test these schema by probing the dark matter halo environments of optically-selected quasars as a function of $g-i$ optical color by measuring the two-point correlation functions of $\sim$ 0.34 million eBOSS quasars as well as the gravitational deflection of cosmic microwave background photons around $\sim$ 0.66 million XDQSO photometric quasar candidates. We do not detect a trend of halo bias with optical color through either analysis, finding that optically-selected quasars at $0.8 < z < 2.2$ occupy halos of characteristic mass $M_{h}\sim 3\times 10^{12} \ h^{-1} M_{\odot}$ regardless of their color. This result implies that a quasar's large-scale halo environment is not strongly connected to its observed optical color. We also confirm findings of fundamental differences in the radio properties of red and blue quasars by stacking 1.4 GHz FIRST images at their positions, suggesting the observed differences cannot be attributed to orientation. Instead, the differences between red and blue quasars likely arise on nuclear-galactic scales, perhaps owing to reddening by a nuclear dusty wind. Finally, we show that optically-selected quasars' halo environments are also independent of their $r-W2$ optical-infrared colors, while previous work has suggested that mid-infrared-selected obscured quasars occupy more massive halos. We discuss implications of this result for models of quasar and galaxy co-evolution. \\

\end{abstract}

%\keywords{Quasars --- 
%Gravitational Lensing}


\section{Introduction} \label{sec:intro}

Quasars are the most luminous class of active galactic nuclei (AGN), objects powered by accretion of interstellar material onto supermassive black holes near the centers of galaxies \citep{1964ApJ...140..796S, 1969Natur.223..690L}. Since their discovery, AGN and quasars have been observationally classified into a taxonomy including many species \citep[e.g.,][]{2017A&ARv..25....2P}. However, a comprehensive physical picture has yet to emerge for many of these variations. In particular, the origin of ``red'' quasars which exhibit redder continuum spectra in the optical through mid-infrared regime than their more numerous blue counterparts remains unexplained. Unified models of AGN \citep{1993ARA&A..31..473A, 1995PASP..107..803U, 2015ARA&A..53..365N} attempt to explain the gamut of AGN varieties in terms of a few parameters intrinsic to the system, stipulating that the array of observed AGN classes are inherently similar yet exhibit differing spectral energy distributions (SEDs) due to chance alignment of the accretion disk along the line-of-sight. This model would attribute red quasars' color to extinction arising from a moderate viewing angle of the dusty ``torus'' \citep[e.g.,][]{2002ApJ...564L..65W, 2013MNRAS.432.2150R}. Alternatively, evolutionary models suggest that AGN may vary in their observed properties over the course of their activity through interaction with their broader environments. One such model of interest in explaining quasars' colors links quasar activity and star formation in a feedback-driven co-evolutionary scheme, which may produce quasars reddened by their host galaxies during merger-triggered starburst events \citep[e.g.,][]{1988ApJ...325...74S, 2001ApJ...555..719C, 2005ApJ...630..705H, 2006ApJS..163....1H, 2008ApJS..175..356H, 2008ApJ...674...80U, 2012NewAR..56...93A, 2012ApJ...757...51G, 2015ApJ...806..218G, 2015MNRAS.447.3368B, 2018ARA&A..56..625H, 2019MNRAS.488.4126P}. A powerful method to distinguish between these schema is to test whether certain classes of AGN exhibit differences across a parameter which cannot feasibly be connected to orientation. Thus, this work aims to probe the nature of optically-selected red and blue quasars by estimating the characteristic dark matter halo mass that each class resides within. 

Despite many investigations \citep[e.g.,][]{1995Natur.375..469W, 2002ApJ...564L..65W, 2003AJ....126.1131R, 2004AJ....128.1112H, 2015AJ....149..203K, 2018A&A...610A..31K, 2019MNRAS.488.3109K, 2021A&A...649A.102C}, the primary driver of optical quasar color has yet to be conclusively determined. The recent work of \citet{2019MNRAS.488.3109K}, \citet{2020MNRAS.494.3061R}, and \citet{2020MNRAS.494.4802F} has demonstrated that optically red quasars exhibit an excess of radio-emission compared to their blue counterparts in a manner that contradicts a purely orientation-based explanation of their colors. Specifically, this work has found that red quasars display higher radio detection fractions, driven primarily by compact radio sources near the radio-quiet/radio-loud threshold. Crucially, this trend is the opposite of that expected if quasars' colors are dominantly determined by their orientation with the observer's line-of-sight, as jets emanating from an edge-on reddened accretion disk should appear more extended and less bright owing to a lack of relativistic beaming toward the observer. Motivated by these results, we set out to test whether optically red and blue quasars display fundamental differences across other properties, particularly their surrounding large-scale structure (LSS). 

As quasars form in overdense regions in the Universe, they are ``biased'' tracers of the underlying matter distribution. In this study, we estimate how red and blue quasars selected with the Sloan Digital Sky Survey (SDSS) sample the matter distribution by measuring both their two-point correlation functions as well as the gravitational lensing of the cosmic microwave background (CMB) induced by their host dark matter halos. This bias can be interpreted with the halo model \citep[e.g.,][]{2000MNRAS.318..203S, 2002PhR...372....1C} in a $\Lambda$-Cold Dark Matter ($\Lambda$-CDM) framework to estimate the characteristic mass of the dark matter halos in which red and blue quasars reside
\citep[e.g.,][]{1999MNRAS.308..119S, 2010ApJ...724..878T}. 

We do not detect any trends of host halo mass with optical quasar color through either method, finding that optically-selected quasars occupy similar dark matter halo environments of $M_{h}\sim 3\times 10^{12} \ h^{-1} M_{\odot}$ across the optical color spectrum. We also confirm fundamental differences in the radio properties between red and blue quasars by performing a stacking analysis of FIRST data. Taken together, these analyses suggest that optically red quasars' colors and enhanced incidence of radio emission are not primarily linked with their surrounding LSS nor their torus' orientation with the line-of-sight. Instead, red quasars' colors likely stem from nuclear-galactic scale processes, perhaps arising from obscuration by a nuclear dusty wind launched by the quasar system itself \citep[e.g.,][]{2000ApJ...545...63E, 2021A&A...649A.102C, 2021MNRAS.505.5283R}.

Throughout this work, we adopt a ``Planck 2018'' CMB+BAO $\Lambda$-CDM concordance cosmology \citep{2020A&A...641A...6P}, with $h = H_0/100 \ \mathrm{km \ s}^{-1} \mathrm{Mpc}^{-1} = 0.6766$, $\Omega_{m} = 0.3111$, $\Omega_{\Lambda} = 0.6888$, $\sigma_{8} = 0.8102$, and $n_{s} = 0.9665$.

\section{Data} \label{sec:data}

\subsection{Quasar Samples}
This work aims to probe the dark matter halo environments of quasars as a function of color through two independent measurements, the two-point correlation function and the gravitational lensing of the CMB. However, the lensing signal sourced from typical individual quasar host halos is orders of magnitude below the noise level of current measurements with Planck, meaning that enormous samples of quasars are required to derive significant results. For the purposes of this work, more sources are required than the largest spectroscopic quasar samples available to date. We therefore elect to use similar but distinct samples for the two analyses. In particular, we use the latest Extended Baryon Oscillation Spectroscopic Survey \citep[eBOSS,][]{2016AJ....151...44D} LSS catalogs for the correlation function measurements and the larger XDQSOz photometric quasar candidate catalog \citep{2015MNRAS.452.3124D} for the lensing analysis. These samples were both optically-selected with SDSS imaging data and have similar color distributions, magnitude limits, and redshift distributions (Figure \ref{fig:redblue}). The results from the two independent analyses and samples can thus be approximately compared. We further describe these samples in the following text, and their properties are summarized in Table \ref{tab:samples}.

\begin{deluxetable}{cccc}
\tablehead{\colhead{Sample} & \colhead{Type} & \colhead{$N_{\mathrm{QSO}}$} & \colhead{Analysis}}
\startdata
XDQSO & Photometric & 656,899 & CMB Lensing \\ 
eBOSS & Spectroscopic & 343,708 & Correlation Functions\\
\enddata
\caption{A summary of the two quasar samples used in this work, including the sample name, selection method, source counts, and use case.}\label{tab:samples}
\end{deluxetable}

\subsubsection{Quasar Sample For Lensing Analysis} \label{sec:parsample}

As large samples of quasars are required to yield statistically significant lensing measurements, we begin with the photometric ``XDQSOz'' catalog \citep{2015MNRAS.452.3124D}, which contains 5,537,436 optically-selected quasar candidates, or 3,874,639 quasars weighted by probability. This catalog was constructed using extreme deconvolution \citep[XD;][]{2011ApJ...729..141B, 2012ApJ...749...41B} to model spectroscopically confirmed quasars in flux-redshift space utilizing GALEX ultraviolet through WISE mid-infrared photometry, while optimally incorporating photometric uncertainties and non-detections. This model was then used to calculate quasar probabilities and photometric redshifts for all point sources in the 8th data release of the SDSS (DR8). The XDQSOz catalog contains all point sources with a quasar probability $P_{\mathrm{QSO}} > 0.2$.

We make several enhancements and quality cuts to this photometric catalog to ensure its usefulness in a statistical study of quasars. First, we match the XDQSOz catalog to the ``DR16Q'' catalog of spectroscopically confirmed quasars \citep{2020ApJS..250....8L}, which contains $\sim 750,000$ bona fide quasars, greater than seven times more objects than appeared in the spectroscopic sample originally used to train the XDQSOz algorithm. We replace the photometric redshifts in the XDQSOz catalog with accurate spectroscopic redshifts and update probabilities to unity for the $\sim 700,000$ matches to spectroscopically confirmed quasars. We also update quasar probabilities to zero for the objects which were targeted by SDSS spectroscopic campaigns and subsequently confirmed not to be quasars. Next, we restrict the catalog by applying the \textbf{good} flag in the XDQSOz catalog which mimics cuts used by the BOSS to remove sources with unreliable photometry. This removes approximately half of the sources. We further require the photometric redshift probability density function to contain only one peak, indicating that the photometric redshift should be reliable. This cut removes an additional quarter of the sample. We thus use the peak redshift estimate in all subsequent analysis, where necessary. Finally, we restrict the catalog to those sources which the XDQSOz algorithm deems high-likelihood quasars, with $P_{\mathrm{QSO}} > 0.9$. This sample therefore consists of uniformly-selected photometric quasar candidates which should be $> 90\%$ pure, and is uniformly enhanced by spectroscopic data when available.

Before splitting this sample by color for cross-correlation with CMB lensing measurements, we make two final cuts. The first selects only quasars with a best redshift estimate between $0.8 < z < 2.2$. This is chosen to match the redshift distribution of the eBOSS sample used for correlation function measurements, such that results from the two analyses may be approximately compared. This redshift range also fortuitously overlaps with the CMB lensing ``kernel'' peak at $z \sim 1-2$ \citep[e.g.,][]{2000ApJ...534..533C, 2003ApJ...590..664S}, and avoids the low quasar selection efficiency at $z > 2.5$ due to degeneracy with stellar colors \citep{2002AJ....123.2945R, 2003AJ....126.1131R}. Finally, this redshift cut avoids low-luminosity quasars at low redshift, which may have colors dominated by host-galaxy contamination rather than dust extinction \citep{2019MNRAS.488.3109K, 2021A&A...649A.102C}. We have thus constructed a catalog of 656,899 uniformly-selected high-probability quasar candidates (58\% of which have been spectroscopically confirmed) suitable for statistical study. We refer to this sample as the ``XDQSO sample'' throughout the remainder of this work.

\subsubsection{Samples For Two-Point Correlation Function Analyses}

We also probe the bias of quasars as a function of color with two-point cross-correlation functions. We perform these measurements with data from the completed eBOSS survey \citep{2016AJ....151...44D}. eBOSS was a spectroscopic survey carried out as part of the SDSS to measure the baryonic acoustic oscillations in the correlation functions of biased tracers of the underlying matter distribution, including quasars, luminous red galaxies (LRGs) and emission line galaxies (ELGs). Following the survey's completion, LSS catalogs for each tracer sample were publicly released along with carefully constructed random catalogs which are designed to match each survey's selection function, a necessary component of performing a clustering measurement. In this work, we estimate the halo bias for quasars of a given color by measuring their cross-correlation with the entire eBOSS quasar sample, as well as the cross-correlation with LRGs and ELGs. We thus utilize the publicly available quasar and LRG LSS catalogs \citep{2020MNRAS.498.2354R} along with the ELG LSS catalog \citep{2021MNRAS.500.3254R}. The eBOSS quasar catalog contains 343,708 spectroscopically confirmed quasars which were selected uniformly using XD probabilities \citep{2012ApJ...749...41B} along with an optical-mid-infrared color cut \citep{2015ApJS..221...27M}. We refer the reader to these publications for details of the catalogs' construction.

%\begin{figure}
%    \centering
%    \includegraphics[width=0.45\textwidth]{xdqso_specz_Mi_z}
%    \caption{The absolute $i$ band magnitude \citep[K-corrected to $z=2$,][]{2006AJ....131.2766R} as a function of redshift for the high probability (PQSO > 0.99) quasar candidates enhanced with spectroscopic data when available. The translucent yellow shading denotes our redshift selection ($0.5 < z < 2.5$) and our absolute magnitude cut of $M_{i} (z=2) < -25$.}
%    \label{fig:MI}
%\end{figure}

\begin{figure*}
    \centering
    \includegraphics[width=0.95\textwidth]{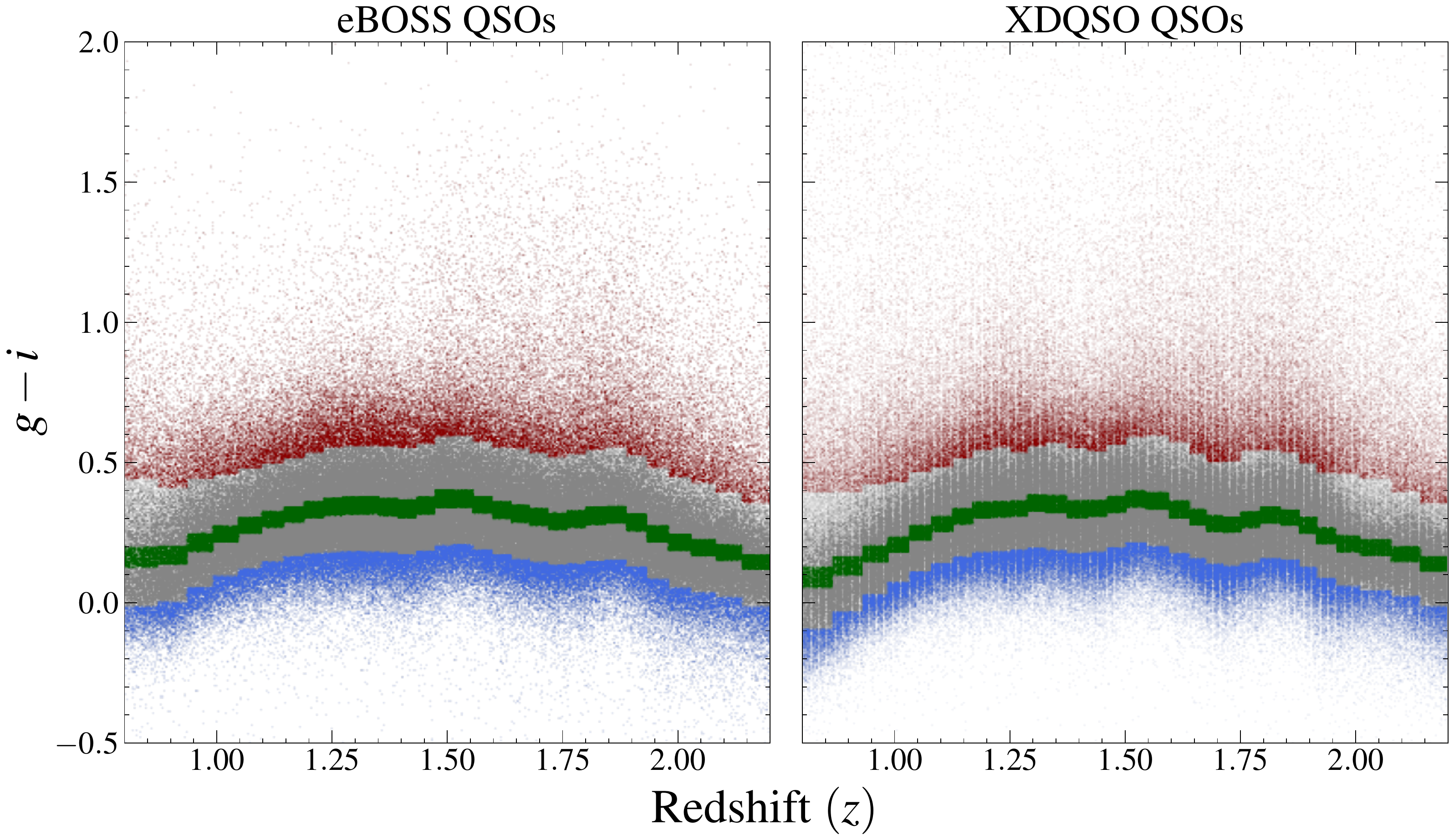}
    \caption{The $g - i$ distribution (corrected for Galactic extinction) for the spectroscopic eBOSS quasars (left panel) and photometric XDQSO quasars (right panel) as a function of redshift. The redshift-evolving color binning scheme described in Section \ref{sec:binning} is demonstrated by highlighting the quasars with $g-i$ color indices belonging to the highest, middle, and lowest 14.3\% quantiles with red, green, and blue, respectively. This demonstrates that the spectroscopic and photometric samples have similar color and redshift distributions and can thus both be used to study the dependence of halo bias on optical quasar color at $0.8 < z < 2.2$.}
    \label{fig:redblue}
\end{figure*}

\subsection{Planck Lensing Convergence Map}
To estimate the average gravitational deflection of CMB photons induced by a given sample of quasar host halos, we utilize the 2018 release of the Planck lensing convergence map\footnote{The lensing products used throughout this work can be accessed at \url{https://pla.esac.esa.int/}} \citep{2020A&A...641A...8P}. We adopt the minimum-variance (MV) estimate combining both temperature and polarization data. We do not use the joint reconstruction with CMB and cosmic infrared background (CIB) data, as our sample of high-redshift quasars may contribute to the CIB and contaminate the lensing estimate. We also make use of the simulated lensing noise maps to estimate uncertainties on our measurements.

To recover signal at the scales of interest, we smooth the map with a 15\arcmin\ FWHM Gaussian beam which removes the highest $l$-modes containing virtually no signal. We also apply an inverse top-hat filter to remove noisy $l$-modes of $l < 100$ \citep{2019ApJ...874...85G}. This filtered map can then be ``stacked'' at the positions of quasars in a given sample to probe the average CMB lensing amplitude generated by their host halos.

\subsection{FIRST Data}

Finally, we utilize data from the Faint Images of the Radio Sky at Twenty-Centimeters (FIRST) survey \citep{1995ApJ...450..559B} undertaken with the Jansky Very Large Array (JVLA) to confirm key differences in the radio properties of red and blue quasars. FIRST was a radio survey at 1.4 GHz over $\sim 10^4 \  \mathrm{deg}^2$ of the northern sky to $\sim 0.15$ mJy RMS depth at $\sim 5'$ resolution, which overlaps the SDSS imaging regions and covers $\sim 84 \%$ of the eBOSS quasar footprint. We extract FIRST imaging cutouts at the positions of quasars using {\tt astroquery} \citep{2019AJ....157...98G} to stack and derive median radio-loudness parameters.

\section{Measurements}

\subsection{Binning Quasars by Color} \label{sec:binning}

%To separate the parent sample into subsamples of optical color, we first examine the distribution of $g-i$ color offsets in Figure \ref{fig:redblue} (right-hand panel). These offsets are computed by subtracting model colors of the \citet{2001AJ....122..549V} composite quasar spectrum from observed colors. As in \citet{2001AJ....121.2308R, 2003AJ....126.1131R}, we observe that the quasar color offsets exhibit an approximately Gaussian distribution, though $\sim 10 \%$ of the quasars lie along an extended tail towards the red end of the spectrum. Alike previous studies, the quasars in the `red tail' of our sample exhibit a broken power law optical spectrum, unlike the remainder of the sample, which is well fit by a single spectral index. We thus adopt the nomenclature of deeming these systems ``dust-reddened quasars''. Throughout our following analyses, we find that these dust-reddened quasars deviate from the trends we find in the remainder of the sample. We thus choose to focus our analysis on objects with $\Delta (g-i) < 0.25$, but include the more heavily reddened sample in our results (using different markers). Importantly this means that the quasars we define as `red' below are moderately red in the range of definitions used within the literature. 

%\begin{figure}
 %   \centering
  %  \includegraphics[width=0.45\textwidth]{colorhist.pdf}
   % \caption{The distribution of color offsets for the parent sample. Color offsets are computed using the composite quasar spectrum of \citet{2001AJ....122..549V}. The distribution is Gaussian with a }
   % \label{fig:color_hist}
%\end{figure}

We adopt a similar technique to that used in \citet{2019MNRAS.488.3109K} in order to separate quasars by color in a redshift-dependent manner. We thus first produce a $g - i$ color distribution \citep[corrected for Galactic reddening;][]{1998ApJ...500..525S, 2011ApJ...737..103S} for each sample in 30 bins of redshift containing equal numbers of sources. Within each redshift bin, we then divide the $g - i$ distribution into 7 bins containing an approximately equal number of objects. We use 7 bins rather than the 10 used by \citet{2019MNRAS.488.3109K} in order to ensure significant detections of the CMB lensing in each bin. This redshift-evolving color selection effectively accounts for the shifting of spectral lines in and out of observed bandpasses at various redshifts, and is thus analogous to performing a $k$-correction to the observed colors before binning. The color as a function of redshift for the reddest, bluest, and median bin of each sample is displayed in Figure \ref{fig:redblue}. It is clear that the two samples have similar redshift distributions and color bin definitions, implying that we can approximately compare the results from the two samples and analyses.

As the definition of a ``red'' quasar varies substantially in the literature, it is important to quantify our delineation. We therefore compute a ``color offset'' for each quasar for the purpose of comparison with other results, given as the difference between the observed color (corrected for Galactic extinction) and a template color for a ``typical'' SDSS quasar using the \citet{2001AJ....122..549V} (VB) template:

\begin{equation}
    \Delta (g-i) = (g-i)_{\mathrm{obs}} - (g-i)_{\mathrm{VB}}.
    \label{eq:vdb}
\end{equation}

We use the VB template as the reference color rather than the median of our sample \citep[e.g.,][]{2019MNRAS.488.3109K}, such that the reader may compare our results with other samples without having to reconstruct the samples used here. We will display the median color offset of each color bin in the following results for this purpose.

\subsection{Weighting Scheme}
\label{sec:weights}

\begin{figure}
    \centering
    \includegraphics[width=0.45\textwidth]{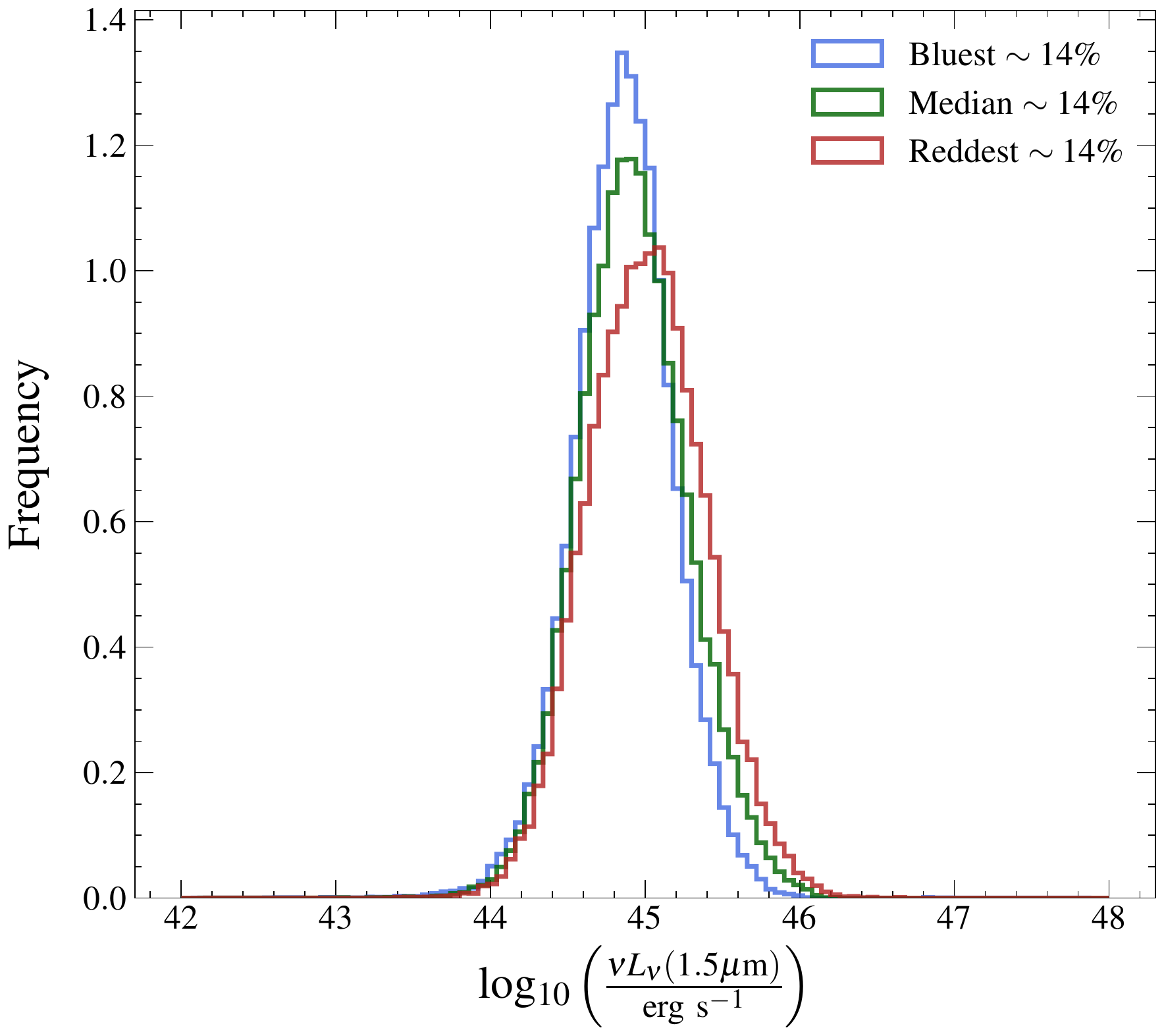}
    \caption{The rest-frame 1.5 $\mu$m luminosity distributions of the reddest, median, and bluest $\sim 14\%$ bins of quasars in the eBOSS sample. We adopt the mid-infrared as a bolometric luminosity indicator as red quasars are likely more affected by dust extinction in the optical. This demonstrates that SDSS quasars with different optical colors exhibit similar but not identical infrared/bolometric luminosity distributions. We assign weights to quasars for all subsequent analyses to control for any effects these small luminosity differences may impart on halo bias measurements.}
    \label{fig:lumhist}
\end{figure}

We develop weights which can be used in both the lensing analysis and the correlation functions to control for any luminosity differences between the different color samples, as halo bias may be correlated with quasar luminosity \citep{2019ApJ...874...85G}, albeit weakly at most \citep{2009ApJ...697.1656S, 2015MNRAS.453.2779E}. We elect to use the mid-infrared to trace the bolometric luminosity of our optically-selected quasars, as this emission should be minimally attenuated by dust. We thus compute rest-frame 1.5 $\mu$m luminosities for each quasar in our sample by interpolating/extrapolating observed 3.4 and 4.6 $\mu$m fluxes from Wide-field Infrared Survey Explorer \citep[WISE;][]{2010AJ....140.1868W} data, assuming a power law spectrum. We use 1.5 $\mu$m emission as the bolometric luminosity tracer rather than the 6 $\mu$m used by \citet[][]{2019MNRAS.488.3109K} so that we do not lose statistical power by requiring that each source is detected in the far less sensitive WISE W3 band. The normalized 1.5 $\mu$m luminosity distributions for each of the color samples are displayed in Figure \ref{fig:lumhist}, demonstrating that the luminosity distributions vary slightly between quasars of differing color. In order to control for these luminosity differences, we first generate normalized 2D distributions of rest-frame 1.5 $\mu$m luminosity and redshift for each color sample. We then compute the minimum value in each luminosity-redshift bin across all of the color samples. Finally, weights are assigned to quasars in each color sample as the ratio of the minimum luminosity-redshift distribution to the distribution of that color sample. This weighting scheme ensures that all of the color samples are matched in redshift and bolometric luminosity. 

For the CMB lensing analysis, we apply a second weighting scheme to account for the fact that some of the sources in the XDQSO sample will be contaminants of stars or galaxies. We thus weight candidates by the quasar probability output by the XDQSO algorithm, $P_{\mathrm{QSO}}$. 

\subsection{Stacking Lensing Map}

Cross-correlations between LSS tracers and CMB lensing convergence maps are typically calculated in Fourier space, but can equivalently be performed in real space, through a ``stacking'' procedure \citep[e.g.,][]{2015PhRvL.114o1302M, 2020ApJ...903L..13M, 2015ApJ...806..247B, 2017NatAs...1..795G, 2019ApJ...874...85G}, which is simpler to implement for samples with complex selection functions. Stacking is the process of averaging over many maps to statistically reveal signals buried beneath the noise of individual maps. Here, we perform a weighted stack of the Planck lensing convergence map at the positions of quasars, where the weights aim to control for differences in properties between samples aside from optical quasar color. For each layer in the stack, we reproject the Planck map using the Lambert azimuthal equal-area projection \citep[with {\tt healpy,}][]{Zonca2019} spanning 4 degrees across centered at the position of each quasar. We then stack by computing a pixel-wise weighted average across all of the projections, ignoring masked pixels.

\begin{figure*}
    \centering
    \includegraphics[width=0.95\textwidth]{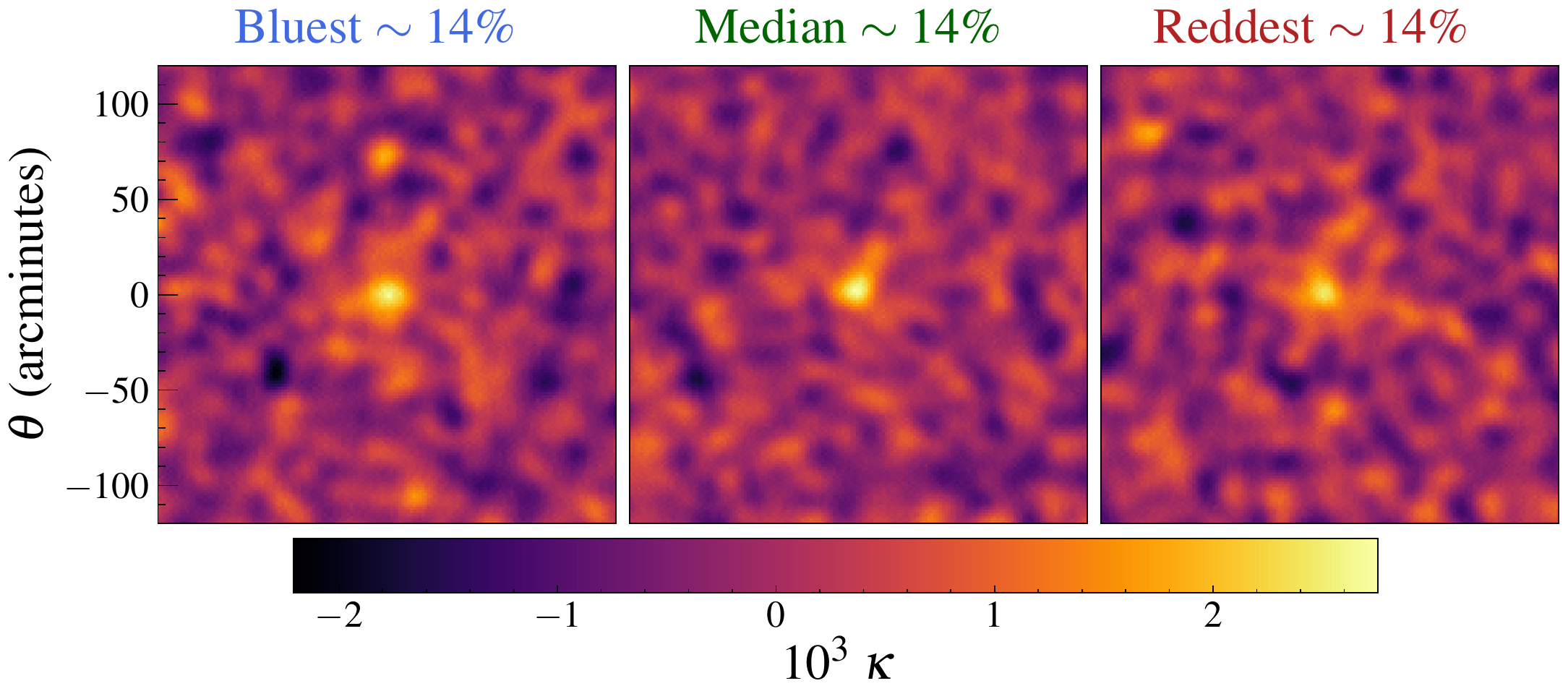}
    \caption{The stacked Planck CMB lensing convergence ($\kappa$) map at the positions of the XDQSO quasars belonging to the bluest, median and reddest bins. Any differences in bolometric luminosity between the samples have been controlled for by stacking with weights described in Section \ref{sec:weights}. No trend of lensing convergence with color is readily apparent, and we quantify the implications for the host halo properties of red and blue quasars in the subsequent analysis. }
    \label{fig:stacks}
\end{figure*}

The result of stacking the Planck lensing convergence map at the positions of the bluest, median, and reddest bins of quasars is shown in Figure \ref{fig:stacks}. As the color samples are constructed to have identical redshift distributions, the peak lensing convergence amplitude ($\kappa$, a unitless surface mass density) directly traces a sample's average halo bias (Eq.\ \ref{eq:avgkappa}). No clear trend of peak $\kappa$ with color is apparent, but we quantify this by fitting these maps in Section \ref{sec:model}.

\subsection{Stacking FIRST Data}

Prompted by the previous finding that red and blue quasars exhibit fundamental differences in their radio properties \citep{2019MNRAS.488.3109K, 2020MNRAS.494.3061R, 2020MNRAS.494.4802F}, we also perform stacks of 1.4 GHz imaging data from the FIRST survey for our quasar samples, where the data is available ($\sim 84 \%$ of the eBOSS footprint). We adopt a median stacking procedure rather than an average stack as per the recommendation of \citet{2007ApJ...654...99W}. We thus retain the $\sim 2\%$ of FIRST-detected sources in our samples, as the median is insensitive to outliers. We also apply the correction for ``CLEAN bias'', which is a systematic underestimation of flux by a factor of 1.4 in stacked FIRST images \citep{2007ApJ...654...99W}. We elect to use the eBOSS spectroscopic quasar sample for this analysis to avoid any contamination by non-quasars and ensure accurate redshift information. It should be noted that we also test stacking radio images at the positions of the XDQSO sample, which gives similar results except for a $\sim 1.5 \sigma$ lower flux in the reddest bin. This may indicate that the reddest candidates in the XDQSO sample suffer from higher contamination fractions or poorer photometric redshifts.

\begin{figure*}
    \centering
    \includegraphics[width=0.95\textwidth]{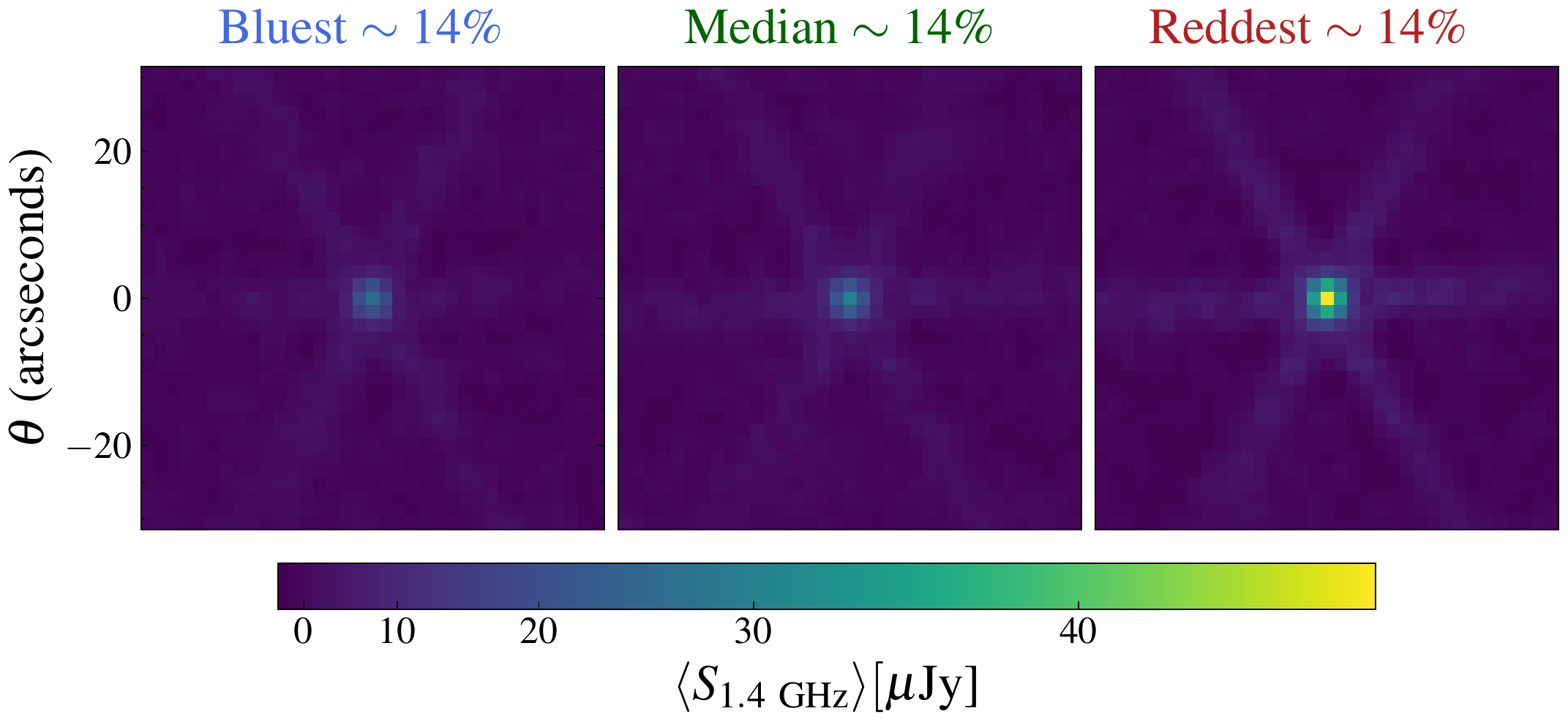}
    \caption{The median stacks of FIRST images at the positions of the bluest, median, and reddest quasars in the eBOSS sample. We have applied a power-law stretch to the color scale to emphasize the ``source'' rather than interferometric artifacts. The flux values have been corrected to account for the ``clean bias'' described by \citet{2007ApJ...654...99W}. It is apparent that the reddest quasars exhibit an excess of 1.4 GHz emission, consistent with previous findings. These stacks have not been controlled for 1.5 $\mu$m luminosity, as we subsequently use them to quantify the median radio-loudness as a function of quasar color. }
    \label{fig:first_stacks}
\end{figure*}

The results of stacking the bluest, median and reddest $\sim 14\%$ of quasars in the eBOSS sample is shown in Figure \ref{fig:first_stacks}, which demonstrates that optically-selected red quasars exhibit an excess of radio-emission, consistent with previous findings \citep{2019MNRAS.488.3109K, 2020MNRAS.494.3061R, 2020MNRAS.494.4802F}. We will quantify this result's effect on the median radio-loudness as a function of quasar color in Section \ref{Results}. We define the radio-loudness as the logarithmic ratio between the 1.4 GHz rest-frame luminosity, and the 1.5 $\mu$m luminosity which is our bolometric luminosity tracer:

\begin{equation}
    R = \log_{10}\left(\frac{\nu L_{\nu}(1.4 \ \mathrm{GHz})}{\nu L_{\nu}(1.5 \ \mu \mathrm{m})}\right).
    \label{eq:loudness}
\end{equation}

We compute median rest-frame luminosities at 1.4 GHz from the stacked flux images using the median redshift of our sample, and by assuming a power law spectrum ($S_{\nu} \propto \nu^{-\alpha}$) with a power law spectral index typical of quasars, $\alpha = 0.5$. \\ \\

\subsection{Two-Point Correlation Functions}
\label{sec:2pcf}

We measure the spatial clustering of eBOSS quasars as a function of color by cross-correlating them with three samples of LSS tracers. First, we measure the cross-correlation of the quasars in each color bin with the entirety of the eBOSS quasar sample. We choose this technique rather than performing autocorrelations of the quasars within each bin in order to mitigate systematics. This is because autocorrelations would require random catalogs representative of the selection function of each color bin, which would be difficult to reconstruct given the nontrivial color-based selection of eBOSS quasars \citep{2015ApJS..221...27M}. Instead, a cross-correlation is dominantly sensitive to the selection function of the ``tracer'' population of the entire eBOSS quasar sample, which has been robustly characterized and may be replicated through adoption of the provided random catalogs and weighting schemes. Next, we measure the cross-correlations of each bin of quasars with both LRGs and ELGs, providing two extra independent probes of the quasars' clustering strength at different effective redshifts. Finally, we must measure autocorrelation functions of the tracer populations of quasars, LRGs and ELGs in order to constrain the absolute bias from the cross-correlations. 

The two-point autocorrelation function $\xi (r)$ is defined as the excess probability above that expected from an unclustered Poisson distribution of finding an object within a volume element $dV$ at a separation $r$ from a randomly chosen initial object in a field with average number density $n$ \citep{1980lssu.book.....P}: 

\begin{equation}
    dP = n [1 + \xi(r)] dV. 
\end{equation}

A cross-correlation function is analogously the excess probability of finding an object in one sample at a separation $r$ from a randomly chosen object in another sample. To estimate the cross-correlations between quasars in a color bin and a given LSS tracer, we adopt the \citet{1983ApJ...267..465D} estimator:

\begin{equation}
    \xi = \frac{Q_{c}T}{Q_{c}R} - 1,
    \label{eq:xcorr}
\end{equation}
where $Q_c T$ represents weighted pair-counts between quasars in a given color bin and the tracer population of either eBOSS quasars, LRGs or ELGs, and $Q_{c} R$ represents weighted counts between the quasars and the randoms representative of the selection function of the tracer population. This estimator is preferred for this analysis as it is independent of the selection function of the quasar sample of interest \citep[e.g.,][]{2007ApJ...654..115C}, which varies with quasar color in a nontrivial fashion. 

We adopt the weights provided within the eBOSS LSS catalogs to ensure the randoms are representative of the selection function of each LSS tracer. These weights incorporate corrections for imaging systematics ($w_{\mathrm{sys}}$), redshift failures ($w_{\mathrm{noz}}$), fiber collisions ($w_{\mathrm{cp}}$) and optimizations for the signal-to-noise at the BAO scale \citep[$w_{\mathrm{FKP}}$,][]{1994ApJ...426...23F}. The weighted pair-counts in Eq.\ \ref{eq:xcorr} are calculated by counting each pair by the product of their weights, where we have weighted the LSS tracers and randoms by $w_{\mathrm{sys}}$, $w_{\mathrm{noz}}$, and $w_{\mathrm{cp}}$, while the quasars in each color bin have been weighted according to our scheme to control for bolometric luminosity differences (Section \ref{sec:weights}).

As we measure redshifts rather than distances to quasars, peculiar motions with respect to the Hubble flow cause the spatial distribution of sources to appear extended or compressed along radial sightlines \citep[e.g.,][]{1987MNRAS.227....1K} -- dubbed ``redshift-space distortions'' (RSDs). Thus, rather than measuring the one-dimensional CF, $\xi(r)$, we estimate the two-dimensional analog: $\xi(r_p, \pi)$, where $r_p$ is the projected separation along a transverse axis, and $\pi$ is the redshift-space separation. As RSDs only affect the line-of-sight component of $\xi$, one can integrate over this axis to derive the projected correlation function $w_p(r_p)$ which is RSD independent \citep{1983ApJ...267..465D}:

\begin{equation}
    w_p(r_p) = 2 \int^{\pi_{\mathrm{max}}}_{0} d \pi \  \xi(r_p, \pi).
    \label{projcf}
\end{equation}

In practice, choosing the integration limit $\pi_{\mathrm{max}}$ is a balance between incorporating a majority of the redshift-space clustering and avoiding excess noise induced by including physically uncorrelated pairs in the statistic. To determine the optimal integration limits, we visually inspect the two-dimensional correlation functions $\xi (r_p, \pi)$ and determine the $\pi$ separations within which most of the signal is contained. We thus use limits of $\pi_{\mathrm{max}}$ of 20, 25, and 15 $h^{-1}$ Mpc for the autocorrelations of eBOSS quasars, LRGs, and ELGs, and use limits of 20, 15, and 15 $h^{-1}$ Mpc for the cross-correlations between quasars and eBOSS quasars, LRGs, and ELGs, respectively.

These cross-correlation measurements simultaneously probe the bias of the quasars in each bin as well as the LSS tracer population. In order to estimate the absolute bias of each set of quasars, we must first estimate the bias of the reference samples of quasars, LRGs, and ELGs. For this, we perform autocorrelation function measurements using the \citet{1993ApJ...412...64L} estimator:

\begin{equation}
    \xi = \frac{DD - 2 DR + RR}{DD}, 
\end{equation}
where $DD$, $DR$, and $RR$ are weighted pair counts normalized by number density as a function of separation for data-data pairs, data-random pairs, and random-random pairs, respectively. We weight both the data and randoms by $w_{\mathrm{sys}}$, $w_{\mathrm{noz}}$, and $w_{\mathrm{cp}}$. When measuring the autocorrelations of ELGs and LRGs, we wish to probe the bias at the same effective redshift as the cross-correlation with quasars. We thus apply additional weights in the autocorrelation given as the overlap between the redshift distribution of the quasars and the galaxy sample:

\begin{equation}
    w_{\mathrm{overlap}} = \sqrt{\frac{dN}{dz}_{\mathrm{QSO}} \frac{dN}{dz}_{\mathrm{GAL}}}.
    \label{eq:overlap}
\end{equation}

We estimate the uncertainty on the measured auto and cross-correlation functions by adopting a variation of the procedure recommended by \citet{2009MNRAS.396...19N}, which found that dividing samples into $N_{\mathrm{sub}}$ subvolumes and resampling with the bootstrap method can approximately reproduce the ``true'' covariance matrix inferred from mock catalogs when the number of resamplings $N_r = 3 N_{\mathrm{sub}}$. We thus divide each quasar and galaxy sample into subvolumes on the sky using HEALPix, and resample as described above. We then perform a second bootstrap resampling of the individual objects selected from the subvolumes \citep[e.g.,][]{2011ApJ...731..117H}. This technique should recover the uncertainty deriving from both the variance of clustering across different regions of the survey footprint and also from the finite counting statistics of the samples. We recalculate the correlation function with each realization, which gives an estimate of the uncertainty to be used for model fitting in the following section.

\section{Modeling}
\label{sec:model}

\subsection{Modeling the Lensing Profile}

In order to estimate the characteristic halo bias for a sample of quasars given a stacked lensing map, we must be able to model the average lensing convergence signal expected from a sample of idealized dark matter halos. We adopt a similar procedure as described by \citet{2017NatAs...1..795G, 2019ApJ...874...85G}, which is summarized here. 

The lensing convergence profile expected from a single spherically symmetric mass distribution such as a dark matter halo (the ``one-halo term'') is given by:

\begin{equation}
    \kappa_{1}(\theta, z) = \frac{\Sigma(\theta)}{\Sigma_{crit}(z)},
    \label{eq:1halo}
\end{equation}
where the numerator is the mass density profile $\rho(r)$ of the halo projected onto the plane of the sky:

\begin{equation}
    \Sigma(R = \theta D_{OL}) = 2 \int^{\infty}_{R} dr \  \frac{r \rho(r)}{\sqrt{r^2-R^2}},
    \label{eq:projection}
\end{equation}
and the denominator is the critical surface mass density for lensing:

\begin{equation}
    \Sigma_{crit}(z) = \frac{c^2}{4 \pi G} \frac{D_{OS}(z)}{D_{LS}(z) D_{OL}(z)},
\end{equation}
where $D_{OS}$, $D_{LS}$, and $D_{OL}$ are the angular diameter distances between the observer and source, the lensing system and the source, and the observer and the lens, respectively. The source in this case is the CMB emission at $z=1100$.

We assume the widely used `NFW' \citep{1997ApJ...490..493N} model for the density profile of a single dark matter halo applicable to Eq.\ \ref{eq:projection}, which is given by:

\begin{equation}
    \rho(r) = \frac{\rho_s}{\left(\frac{r}{r_{s}}\right)\left(1+\frac{r}{r_{s}}\right)^{2}}.
    \label{eq:NFW}
\end{equation}

We adopt a standard scale radius $r_s = r_{200}/c$, where $r_{200}$ is the radius at which the enclosed halo is overdense with respect to the universe's critical density by a factor of 200, and $c$ is the ``concentration'' parameter. We assume the mass-concentration-redshift relation of \citet{2016MNRAS.460.1214L}, and convert this to a bias-concentration relation with the \citet{2010ApJ...724..878T} model. We thus characterize the profile of Eq.\ \ref{eq:NFW} in terms of halo bias and redshift rather than $\rho_s$ and $r_s$.

The observed lensing convergence profile around a halo will be the superposition of the one-halo term and the ``two-halo term'', the lensing due to correlated LSS, derived using the \citet{1953ApJ...117..134L} approximation \citep[e.g.,][]{2011MNRAS.414.1851O}:

\begin{equation}
    \kappa_2(\theta, z) = \frac{\overbar{\rho_m}(z)}{(1+z)^3\Sigma_{crit}(z) D_{A}^2(z)} b_h \int \frac{l dl}{2 \pi} J_0(l \theta) P_m(k, z),
\end{equation}
where $\overbar{\rho_m}(z)$ is the average matter density of the universe at redshift $z$, $J_{0}$ is the zeroth-order Bessel function, $b_h$ is the linear halo bias, $P_m$ is the matter power spectrum, and $k = l/[(1+z)D_A]$. We specify the linear halo bias in terms of the spherical overdensity halo mass and redshift by adopting the model of \citet{2010ApJ...724..878T}.

The total predicted lensing convergence amplitude from a population of dark matter halos and their surrounding LSS is given by the sum of the one and two-halo terms, averaged over the normalized redshift distribution of the lenses, $\frac{dn}{dz}$:
\begin{equation}
    \langle \kappa \rangle  = \int dz \  (\kappa_1 + \kappa_2) \frac{dn}{dz}.
    \label{eq:avgkappa}
\end{equation}

The model of Eqs.\ \ref{eq:1halo}-\ref{eq:avgkappa} produces the expected $\kappa$ profile for a given redshift distribution, and takes only the halo bias as a free parameter. To compare this model to our stacked data, we first filter the model in the same manner as the data, smoothing with a $15$\arcmin\ Gaussian and removing Fourier modes $l < 100$. We then project the model onto the same grid as the data and measure both the data and the model profile by azimuthally averaging over radial bins. We then vary the halo bias in the model to optimize the fit as follows.

The smoothing applied to the data creates covariance between angular bins. We thus construct a covariance matrix by computing:

\begin{equation}
    C_{ij} = \frac{1}{N-1}\sum_{k=0}^{N} (\kappa_{i}^k - \overbar{\kappa_i})(\kappa_{j}^k - \overbar{\kappa_j}), 
    \label{covar}
\end{equation}
where $i, j$ are labels for the angular bins and each $k$ represents a different realization of the stacked noise map. We construct these realizations by filtering and stacking 60 simulated noise maps provided by Planck in the same manner as the data maps.

We can now estimate a best fitting model by optimizing the likelihood function:

\begin{equation}
    \chi^2 = (\kappa_{\mathrm{obs}} - \kappa_{\mathrm{model}})^T C_{ij}^{-1} (\kappa_{\mathrm{obs}} - \kappa_{\mathrm{model}}).
    \label{chisquare}
\end{equation}

\begin{figure}
    \centering
    \includegraphics[width=0.45\textwidth]{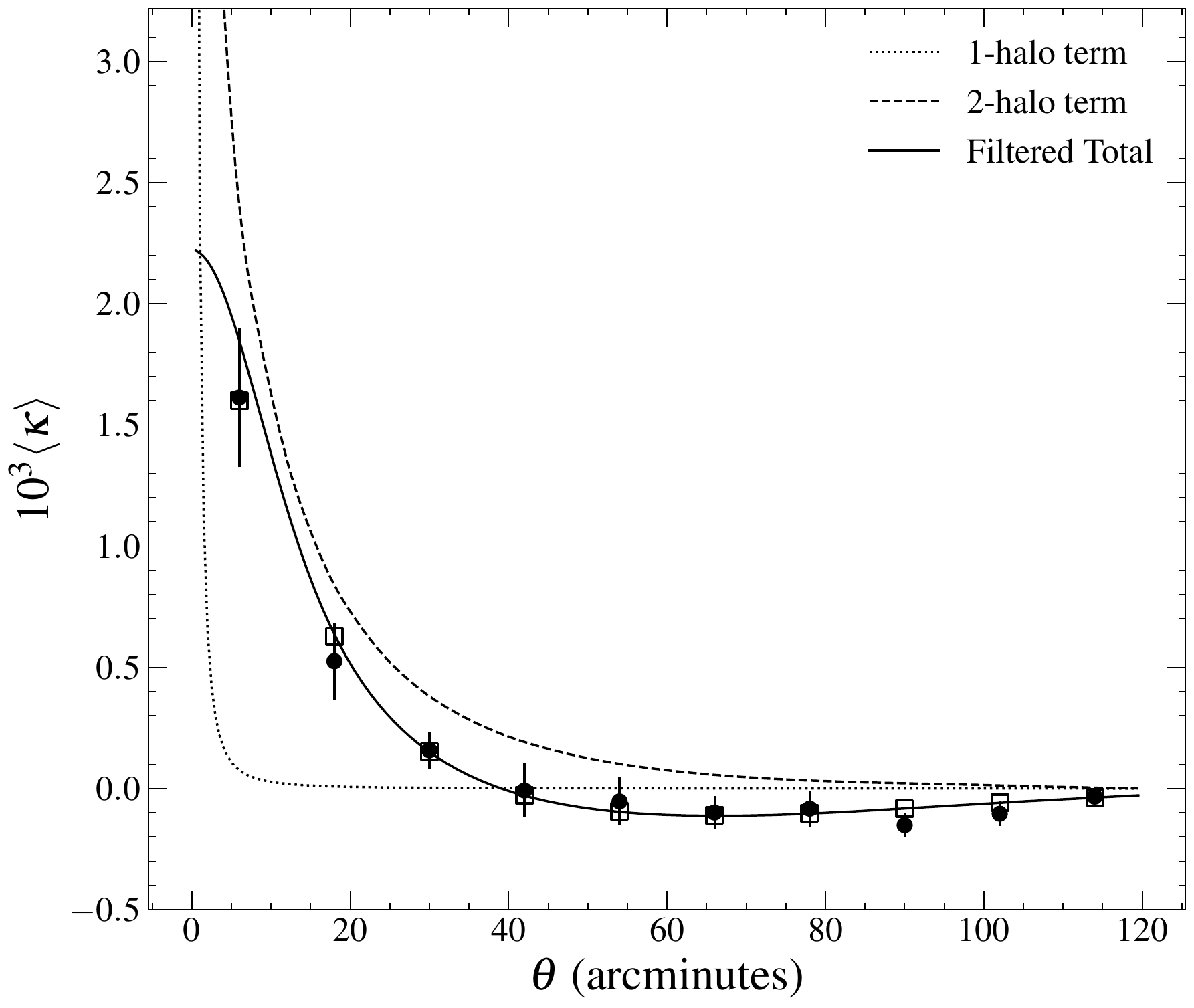}
    \caption{The measured lensing convergence profile for the stack of the median $\sim 14\%$ of XDQSO quasars (Figure \ref{fig:stacks}, middle panel) and the best fit model. The circular points represent average $\kappa$ values measured in annular bins of the stacked lensing map. The solid line represents the best fit model comprising the sum of the one and two-halo terms, after filtering. The open squares represent the value of the best fit model ``measured'' in the same bins as the data.}
    \label{fig:profile}
\end{figure}

To evaluate the uncertainty on the best fit halo bias, we re-perform the fit on 60 maps generated by stacking bootstrap resamplings of the relevant quasar sample. In Figure \ref{fig:profile}, we illustrate an example of the result of this procedure by displaying the measured stacked lensing convergence profile for the middle color bin of the XDQSO quasar sample, along with the best fit model.

\subsection{Modeling Correlation Functions}

We also model the dark matter projected correlation function in our cosmology to fit the observed correlation functions and derive bias values. We thus first generate matter power spectra on a grid across the redshift span of each sample using {\tt CAMB} \citep{2000ApJ...538..473L}, then Fourier transform the power spectra to derive correlation functions. Next, we project the real space correlation functions $\xi(r)_{dm}$ to projected correlation functions $w_{p} (r_{p})_{dm}$ with the Abel transform:

\begin{equation}
    w_p(r_p)_{dm} = 2 \int_{r_p}^{\infty} dr \  \frac{r \xi(r)_{dm}}{\sqrt{r^2 - r_p^2}}.
\end{equation}
Finally, we perform a weighted average of the dark matter correlation functions at various redshifts, weighted by the redshift distribution of the sample (or the redshift overlap in the case of a cross-correlation, Eq.\ \ref{eq:overlap}). The resulting correlation function is averaged in the same radial bins as the data and used to fit for a bias value. As we model only the two-halo term of the correlation function for the purposes of this study, we perform all fits on scales of $5 < r_{p} < 25 \  h^{-1}$ Mpc, a regime which is dominated by the two-halo term and is governed by linear structure growth.  

In order to estimate the absolute bias of a given set of quasars from a cross-correlation, one must have an estimate of the bias of the tracer population which is being correlated against. We thus first fit the autocorrelation functions of eBOSS quasars, LRGs, and ELGs to estimate their absolute bias.
First, we compute the average of correlation functions at different redshifts, weighted by the amplitude of the normalized redshift distribution $dn/dz$ of the corresponding tracer sample. The linear bias $b$ of the sample with respect to the underlying matter distribution can then be fit according to:

\begin{equation}
    w_p(r_p) = b^2 w_p(r_p)_{dm}.
\end{equation}

The cross-correlation function between two LSS tracers can be fit by the product of the bias of each tracer with the dark matter correlation function:
\begin{equation}
    w_p(r_p)_{QT} = b_{1} b_{2} w_p(r_p)_{dm}.
\end{equation}

The uncertainties on the clustering measurements derive from bootstrap resampling subvolumes of the data, described in Section \ref{sec:2pcf}. With these realizations, we construct a covariance matrix in an analogous manner to Eqs.\ \ref{covar} \& \ref{chisquare}. However, we find that fitting our data with these matrices give poor results due to noisy cross-terms, a problem noted by similar studies \citep[e.g.,][]{2009ApJ...697.1656S}. Therefore, we use only the diagonal of the covariance matrix in our fitting. We fit each realization for the bias parameter and take the variance of the results as our uncertainty for the bias of the full sample.

\section{Results} \label{Results}
\label{sec:results}

\begin{figure}
    \centering
    \includegraphics[width=0.45\textwidth]{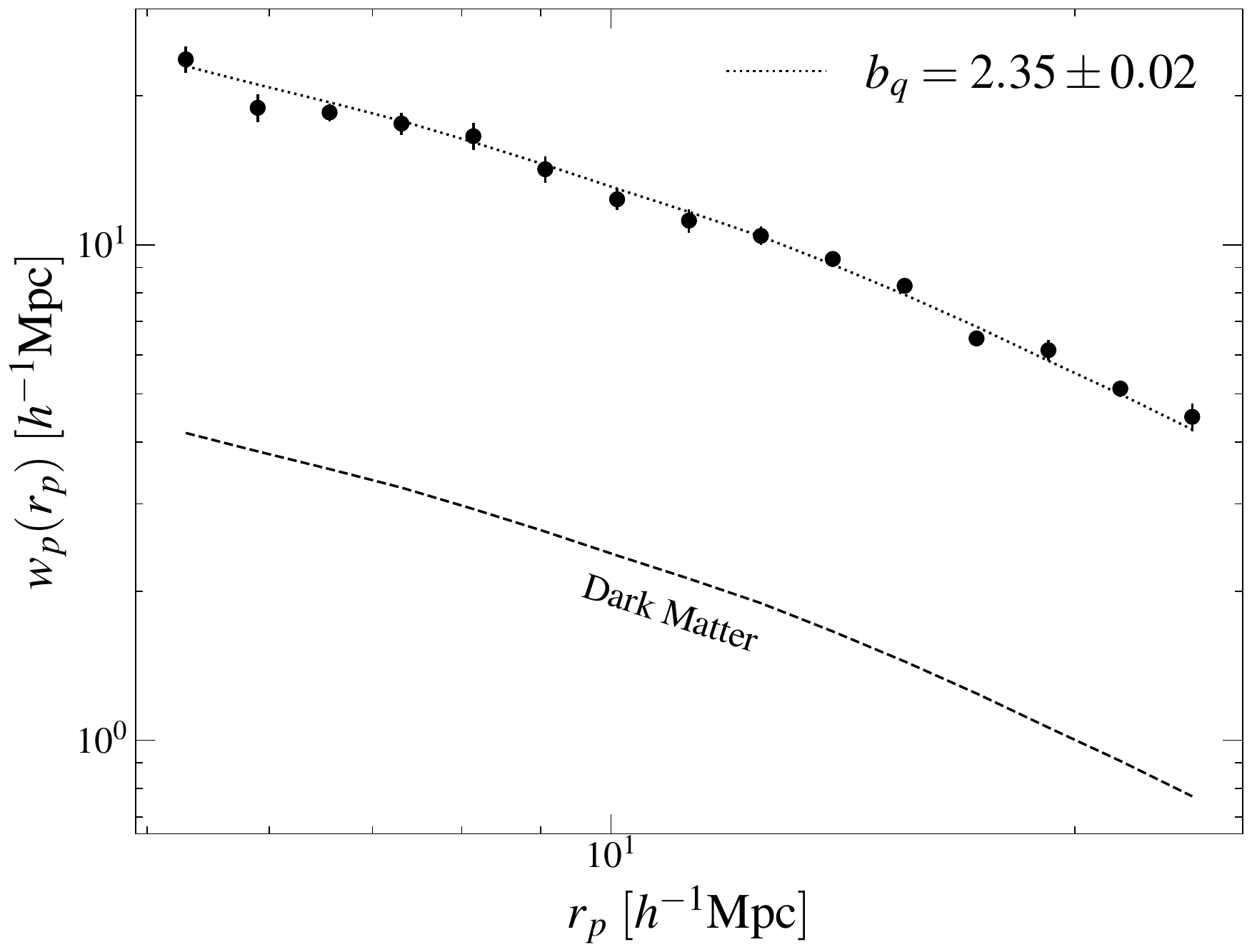}
    \caption{The projected autocorrelation function of eBOSS quasars (black circles), along with the modeled dark matter projected correlation function (dashed line) and the best model fit (dotted line). The dark matter model fits to the autocorrelations of LRGs and ELGs are of similar quality.}
    \label{fig:qso_auto}
\end{figure}

\begin{figure}
    \centering
    \includegraphics[width=0.45\textwidth]{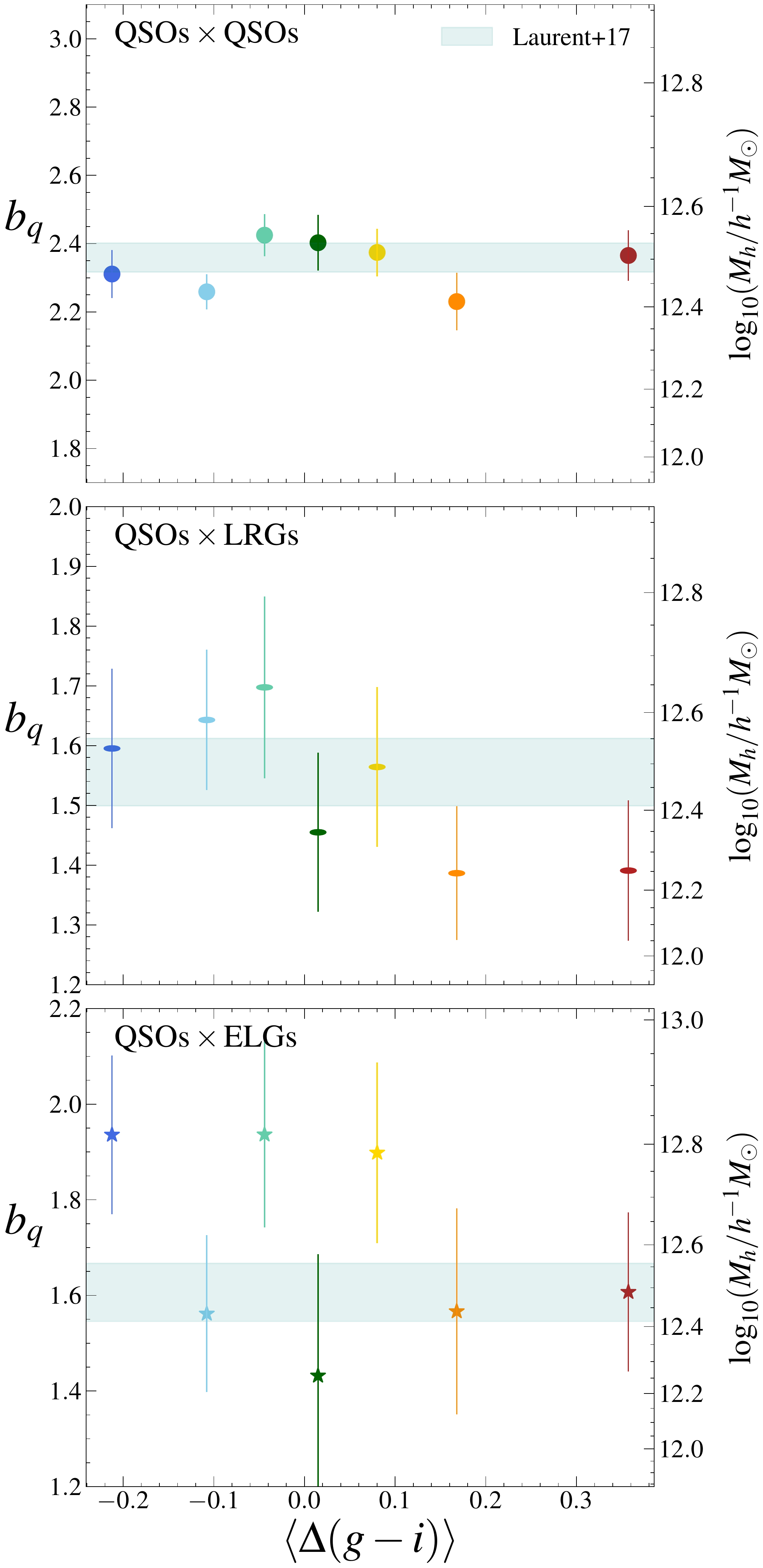}
    \caption{The measured bias as a function of median color offset for the three correlation function analyses. Top panel: the results from the cross-correlation of quasars with the entirety of the eBOSS quasar sample. Middle panel: the cross-correlation of quasars with LRGs. Bottom panel: the cross-correlation of quasars with ELGs. The circles, ellipses and stars show the best-fit bias from cross-correlations with quasars, LRGs, and ELGs. All three panels demonstrate a constant halo bias with quasar color within the uncertainties, appearing to scatter about bias estimates of optically-selected quasars from the literature \citep[shown as teal bands representing 1$\sigma$ bounds from][]{2017JCAP...07..017L}. The secondary ordinate shows characteristic halo masses for the measured biases, demonstrating that quasars occupy similarly massive halos regardless of color and effective redshift.}
    \label{fig:bias}
\end{figure}

In Figure \ref{fig:qso_auto}, we display the autocorrelation function of eBOSS quasars and the best fitting model, with a bias value of $b_{q} = 2.35 \pm 0.02$. This is in excellent agreement with the finding from \citet{2017JCAP...07..017L}, which used a preliminary subset of the eBOSS quasar sample to estimate the evolution of quasar bias as a function of redshift, which gives $b_q = 2.36 \pm 0.04$ for our sample's redshift distribution. It is also fully consistent with the parameterization as probed by the Two Degree Field Quasar Redshift (2QZ) survey given in \citet{2005MNRAS.356..415C}, which predicts a bias of $b_q = 2.39 \pm 0.31$. The above measured bias value for the entire eBOSS quasar sample will be combined with the cross-correlation measurements to derive the absolute bias of each color sample. We estimate the bias of LRGs and ELGs in the same manner, which can then be used to probe the absolute bias of quasars through their cross-correlations.

We display the results of all three cross-correlation analyses, including the best fit bias parameter and corresponding characteristic halo mass for each color bin of quasars in Figure \ref{fig:bias}, and record the fit parameters in Table \ref{tab:xcorr}. These analyses demonstrate that optically-selected quasars of differing color occupy dark matter halos of similar mass. This result appears to hold across all three independent analyses which probe different effective redshifts. The results of each analysis also scatter around the \citet{2017JCAP...07..017L} prediction for eBOSS quasars of all colors, demonstrating that our results are both self-consistent and consistent with previous findings in the literature.

\begin{figure}
    \centering
    \includegraphics[width=0.45\textwidth]{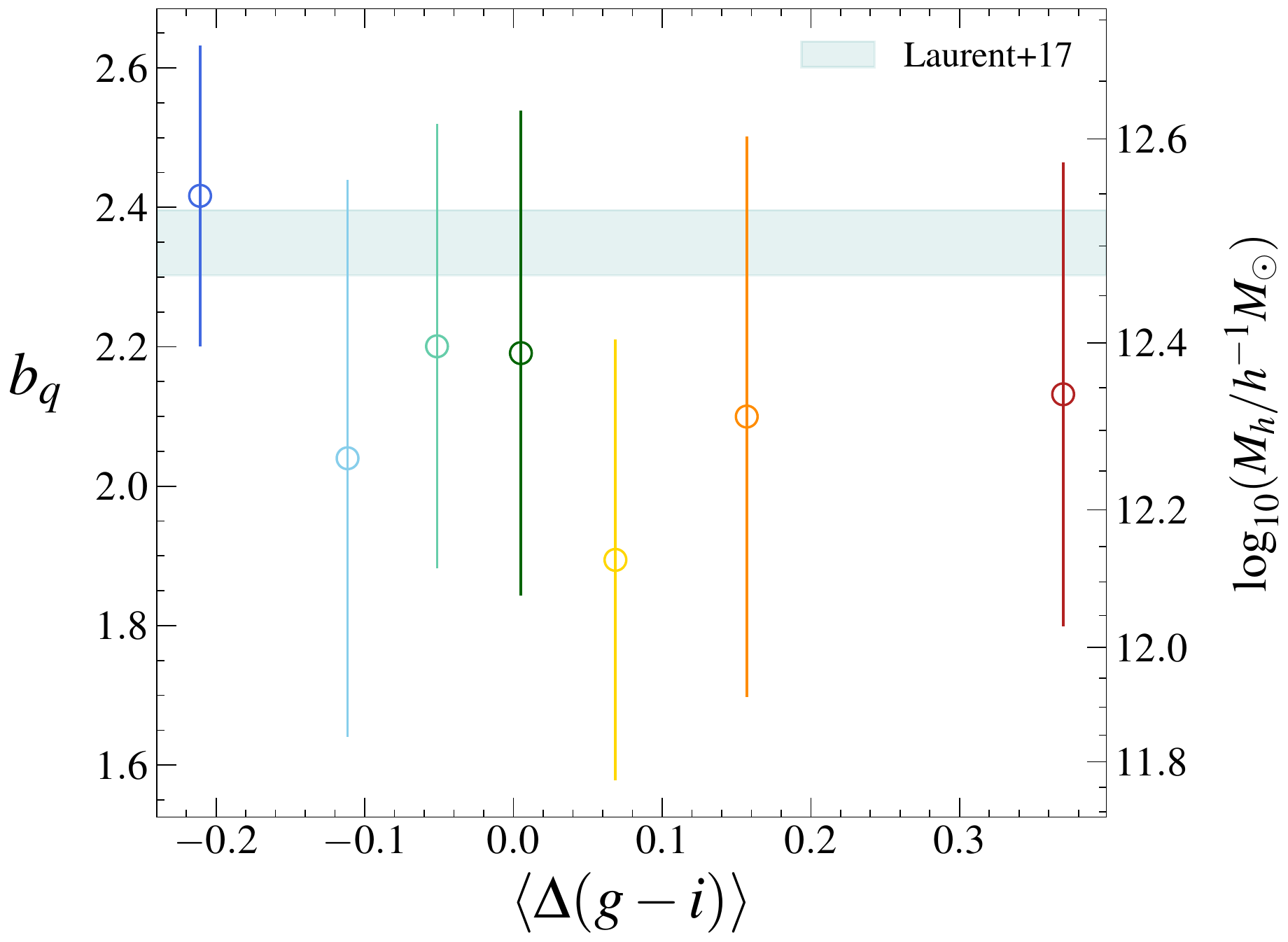}
    \caption{The measured bias (left ordinate) and implied characteristic host halo mass (right ordinate) as a function of median quasar color offset as measured through the Planck CMB lensing analysis of XDQSO photometric quasar candidates. Similarly to the results of the cross-correlation functions of eBOSS quasars (Figure \ref{fig:bias}), this analysis does not reveal any trend of halo bias with quasar color.}
    \label{fig:lensing_bias}
\end{figure}

\begin{figure}
    \centering
    \includegraphics[width=0.43\textwidth]{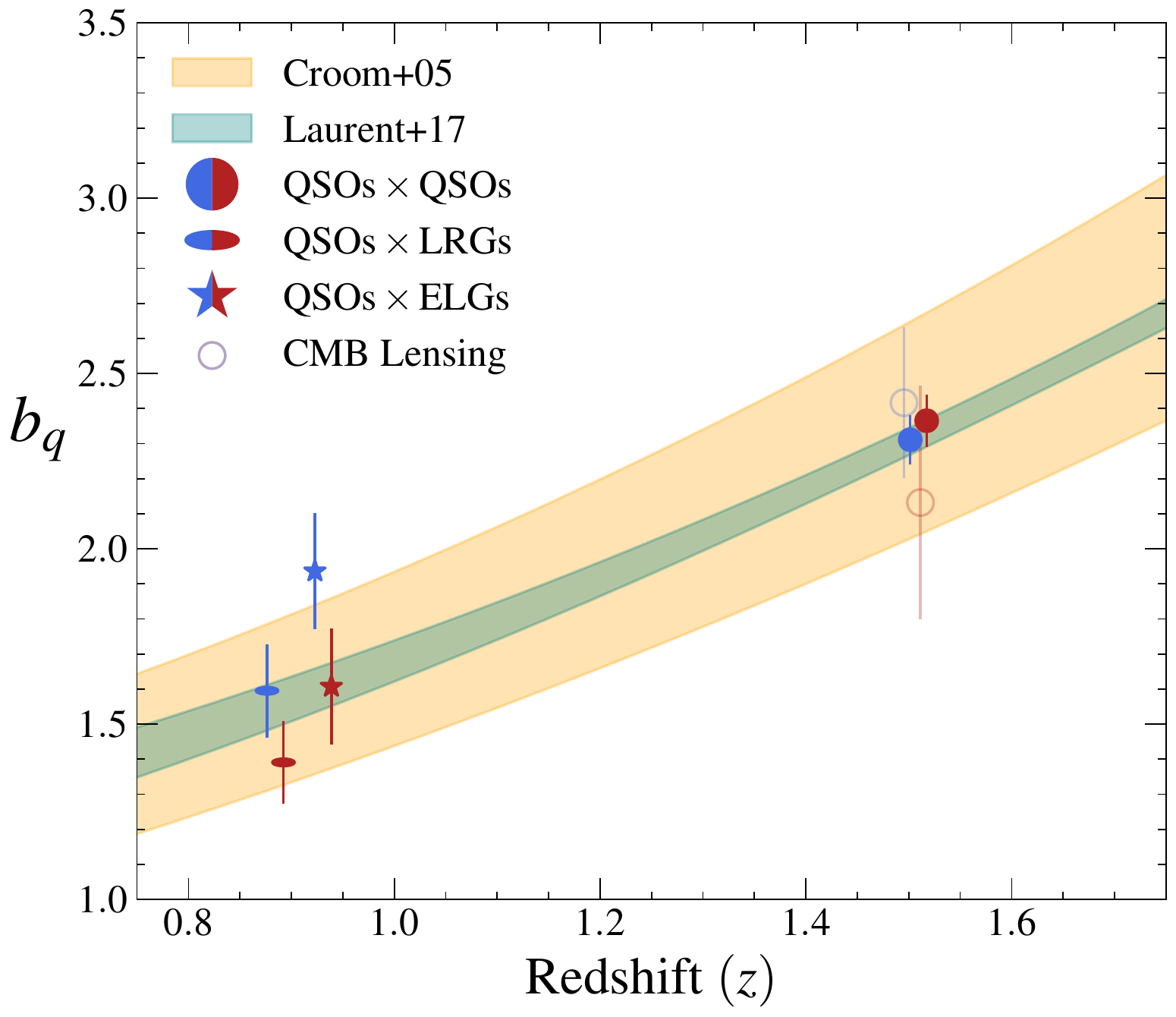}
    \caption{The bias of the reddest $\sim 14\%$ and bluest $\sim 14\%$ of quasars as probed by all of the above analyses at different effective redshifts. Circular markers, stars, and ellipses show results from cross-correlations with eBOSS quasars, ELGs, and LRGs, respectively. Open transparent circles display the results from the CMB lensing analysis of XDQSO photometric quasar candidates. All markers have been shifted slightly from their true effective redshifts for clarity. The parameterizations of optically-selected quasar bias with redshift from \citet{2005MNRAS.356..415C} and \citet{2017JCAP...07..017L} are shown with their $1 \sigma$ bounds in orange and teal, respectively. The bias of the reddest and bluest quasars we measure in this work are consistent with the parameterizations from the literature, as well as with each other across $0.8 < z < 2.2$. Thus, optically red and blue quasars appear to occupy similar dark matter environments. }
    \label{fig:bias_z}
\end{figure}

By utilizing the independent technique of measuring the CMB lensing around XDQSO quasars, we find results entirely consistent with the cross-correlation analyses. Figure \ref{fig:lensing_bias} displays the bias and halo mass derived from measuring the gravitational lensing signal of quasars' host halos and demonstrates a similar result. We display a summary of all of the above bias measurements in Figure \ref{fig:bias_z}, which shows the measured bias of the bluest and reddest bins of quasars as probed by the cross-correlations and the lensing analysis. It is apparent that our measured values are consistent with those found in the literature for optically-selected quasars. This figure also restates that the reddest and bluest optical quasars sample the underlying matter distribution in a statistically indistinguishable manner at high significance. This is in concordance with the findings of \citet{2009ApJ...697.1656S}, which found no difference in the clustering of quasars of different $g-i$ color, albeit at lower significance.  Thus, the primary driver of optical quasar color is likely unrelated to the large-scale environment, which we further discuss in Section \ref{sec:discussion}.

\begin{table*}
    \parbox{.1\linewidth}{
        \centering
            $\langle \Delta (g-i) \rangle$
            \begin{tabular}{c}
            \hline
            \hline
            \\
            \\
            \hline
                 -0.22  \\
                 -0.11  \\
                 -0.05\\
                 0.02\\
                 0.07\\
                 0.17\\
                 0.36
            \end{tabular}
    }
    \parbox{.3\linewidth}{
      \centering
        (a) QSOs $\times$ QSOs
        \begin{tabular}{cc}
            
            \hline
            \hline
            $b_q$ & $M_h$\\ 
             & $10^{12} \ h^{-1} \ M_{\odot} $\\
            \hline
            2.31$ \pm $0.07 & 2.95$ \pm ^{0.35}_{0.32}$ \\
            2.26$ \pm $0.05 & 2.71$ \pm ^{0.24}_{0.23}$ \\
            2.42$ \pm $0.06 & 3.52$ \pm ^{0.33}_{0.32}$ \\
            2.40$ \pm $0.08 & 3.41$ \pm ^{0.44}_{0.41}$ \\
            2.37$ \pm $0.07 & 3.26$ \pm ^{0.36}_{0.34}$ \\
            2.23$ \pm $0.08 & 2.58$ \pm ^{0.39}_{0.36}$ \\
            2.37$ \pm $0.07 & 3.22$ \pm ^{0.38}_{0.36}$ \\
        \end{tabular}
    }
    \parbox{.3\linewidth}{
      \centering
        (b) QSOs $\times$ LRGs
        \begin{tabular}{cc}
            \hline
            \hline
            $b_q$ & $M_h$\\ 
             & $10^{12} \ h^{-1} \ M_{\odot} $\\
            \hline
            1.60$ \pm $0.13 & 3.40$ \pm ^{1.37}_{1.11}$ \\
            1.64$ \pm $0.12 & 3.86$ \pm ^{1.27}_{1.07}$ \\
            1.70$ \pm $0.15 & 4.42$ \pm ^{1.81}_{1.47}$ \\
            1.45$ \pm $0.13 & 2.23$ \pm ^{1.10}_{0.86}$ \\
            1.56$ \pm $0.13 & 3.11$ \pm ^{1.31}_{1.06}$ \\
            1.39$ \pm $0.11 & 1.76$ \pm ^{0.80}_{0.63}$ \\
            1.39$ \pm $0.12 & 1.79$ \pm ^{0.85}_{0.67}$ \\
        \end{tabular}
    }
    \parbox{.3\linewidth}{
      \centering
        (c) QSOs $\times$ ELGs
        \begin{tabular}{cc}
            \hline
            \hline
            $b_q$ & $M_h$\\ 
             & $10^{12} \ h^{-1} \ M_{\odot} $\\
            \hline
            1.94$ \pm $0.17 & 6.56$ \pm ^{2.29}_{1.93}$ \\
            1.56$ \pm $0.16 & 2.72$ \pm ^{1.46}_{1.12}$ \\
            1.94$ \pm $0.19 & 6.56$ \pm ^{2.72}_{2.22}$ \\
            1.43$ \pm $0.25 & 1.81$ \pm ^{1.99}_{1.20}$ \\
            1.90$ \pm $0.19 & 6.09$ \pm ^{2.55}_{2.07}$ \\
            1.57$ \pm $0.22 & 2.76$ \pm ^{2.00}_{1.41}$ \\
            1.61$ \pm $0.17 & 3.09$ \pm ^{1.58}_{1.22}$
        \end{tabular}
    }%
    \caption{A tabulation of the best dark matter model fit parameters to the cross-correlations of eBOSS quasars with other eBOSS quasars (a), LRGs (b), and ELGs (c), including the halo bias as well as the characteristic halo mass. These results are represented visually in Figure \ref{fig:bias}.}\label{tab:xcorr}
\end{table*}

\begin{figure}
    \centering2
    \includegraphics[width=0.45\textwidth]{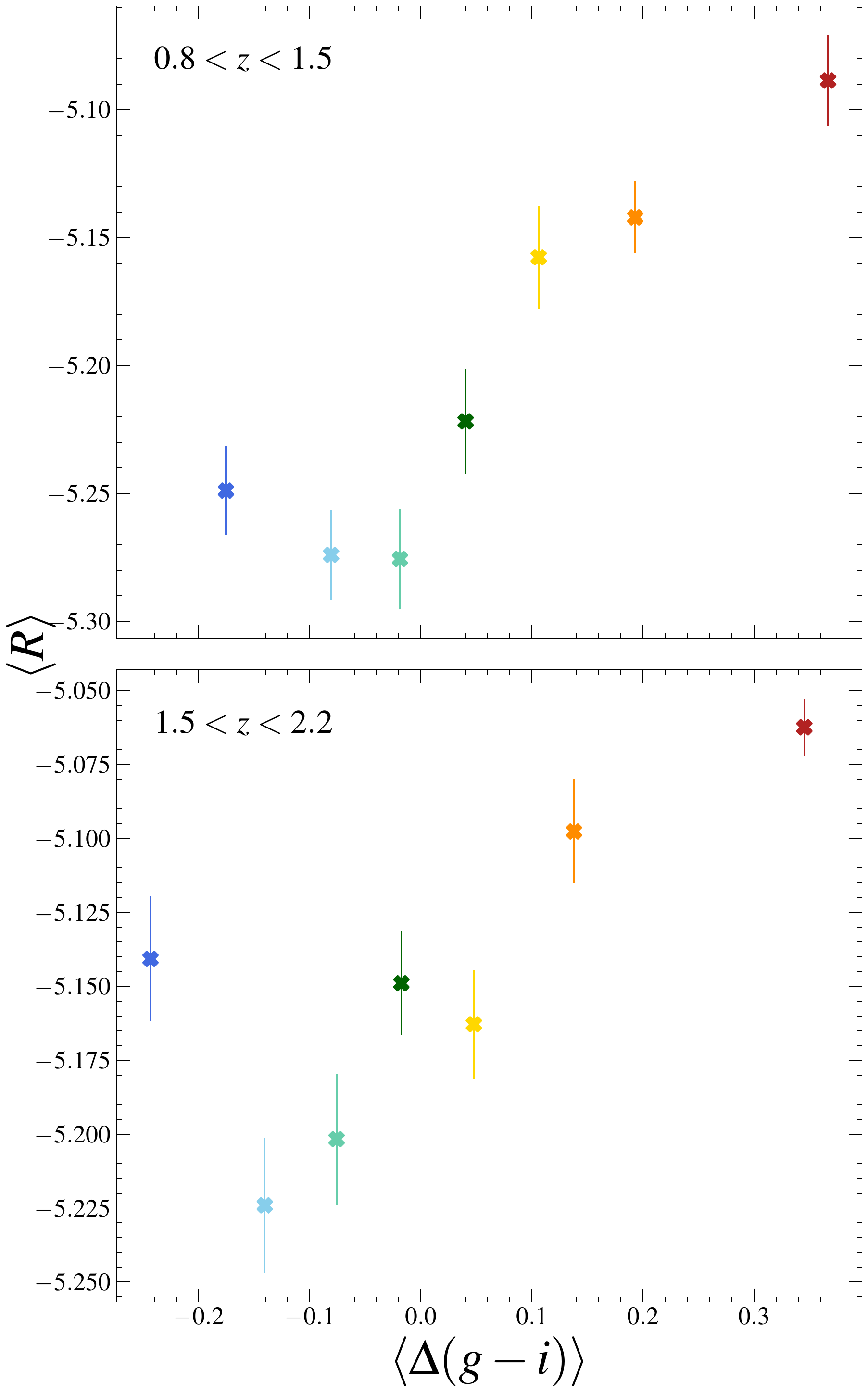}
    \caption{The median radio-loudness (Eq.\ \ref{eq:loudness}) of the eBOSS quasar sample as a function of median color offset \citep[from the ][template, Eq.\ \ref{eq:vdb}]{2001AJ....122..549V}. We compute this in two redshift bins, $0.8 < z < 1.5$ (top panel) and $1.5 < z < 2.2$ (bottom panel). These values were determined by median stacking FIRST images at the positions of quasars. The median radio-loudness clearly increases as quasars appear redder. The exception is the bluest bin, which may be interpreted as relativistic beaming. These values are all well below the classical radio-loud threshold of $R$ = -4.24, converted from \citet{2019MNRAS.488.3109K} using the Type 1 quasar template of \citet{2017ApJ...849...53H}.}
    \label{fig:loudness}
\end{figure}

In Figure \ref{fig:loudness}, we display the median radio-loudness as a function of quasar color for the eBOSS sample, defined in Eq.\ \ref{eq:loudness}, where the median radio luminosity is derived from stacks of 1.4 GHz FIRST images. A clear relationship appears between quasar color and radio-loudness, similar to the trend found by \citet{2007ApJ...654...99W}. Namely, the radio-loudness increases towards redder systems, except for a possible reversal in the bluest bin. This uptick may be understood as evidence that some proportion of quasars' colors stem from their orientation with the line of sight, as a blue face-on accretion disk with a jet traveling towards the observer will be relativistically beamed and radio-enhanced. However, the bulk trend of increasing radio-loudness in redder quasars is not immediately accounted for by an orientation-based mechanism, and bolsters the finding of excess FIRST detection fractions in optically red quasars \citep{2019MNRAS.488.3109K}. As a check, we also test the FIRST-detection fraction as a function of color, finding that the reddest bin are detected $\sim 2.5$ times more often than quasars belonging to the bluest bin, consistent with the findings of \citep{2019MNRAS.488.3109K, 2020MNRAS.494.4802F, 2020MNRAS.494.3061R}. We conclude that the same differences in radio-properties highlighted by these studies appear in our sample, buttressing the result that orientation is not driving quasars' colors.

\section{Discussion}
\label{sec:discussion}

\subsection{Interpretation}

Robustly interpreting the clustering strength of different samples of quasars requires considering all other properties which might scale with halo bias, including redshift, luminosity, and also host galaxy properties like stellar mass and star formation rate (SFR) \citep{2016ApJ...821...55M}. In this work, we have ensured that the quasar samples of different colors are matched in redshift and 1.5 $\mu$m luminosity by applying weighting schemes. We also note that \citet{2021A&A...649A.102C} recently performed ultraviolet through far-infrared SED modeling on SDSS red and blue quasars to investigate their multiwavelength properties, finding that they have statistically identical distributions of stellar mass and SFR. Thus, we argue that we can interpret the clustering strength as a function of quasar color measured here as being uncontaminated by other differences between the samples.

We do not detect any trends of halo bias with optical quasar color from CMB lensing measurements of XDQSO quasars, nor from cross-correlations of eBOSS quasars with three independent tracers, robustly demonstrating that quasars exhibiting different optical colors reside in similar dark matter halo environments. This finding alone is consistent with the unified model of AGN, which attributes color to the chance inclination of the accretion disk/torus along the line-of-sight and would thus predict no trend of color with halo bias. However, recent studies have found that optically red quasars exhibit an excess of compact radio emission near the radio-quiet/radio-loud threshold, a result in direct tension with the unified model \citep{2019MNRAS.488.3109K, 2020MNRAS.494.3061R, 2020MNRAS.494.4802F}. According to the unified model, an inclined and red quasar would be less likely to exhibit excess radio emission owing to a lack of relativistic beaming toward the observer, and would would appear more extended due to jets propagating perpendicularly to the line-of-sight. In this work, we have confirmed the excess radio-loudness in red quasars through stacking of FIRST data. With a suite of analyses now demonstrating similar findings from a range of different quasar samples and radio datasets, the differences between red and blue quasars cannot be primarily attributed to either orientation or large-scale environment.

We therefore suggest that the differences between red and blue quasars are generated on nuclear-galactic scales (roughly between the black hole's gravitational sphere of influence and the galaxy's extent, $10 \lesssim r \lesssim 10^4$ pc). This result is consistent with the finding that the excess radio-emission from red quasars is primarily driven by compact radio sources rather than extended classical radio-loud sources \citep{2019MNRAS.488.3109K}. \citet{2021MNRAS.505.5283R} recently studied a sample of red and blue quasars with high-resolution e-MERLIN imaging, confirming that red quasars exhibit excess radio emission only when the radio source is contained within the extent of the host galaxy. This interpretation is also supported by the work of \citet{2021A&A...649A.102C}, which found that SDSS red/blue quasars exhibit strikingly similar SEDs from the far-infrared to the ultraviolet aside from an excess of $2-5 \ \mu$m emission correlated with optical reddening. This excess was found to coincide with broad [OIII] lines associated with outflows, suggesting that winds of hot dust on nuclear scales may be the dominant source of optical quasar reddening. This could be consistent with the discovery of dust structures along the polar axis of some local AGN \citep{2013ApJ...771...87H, 2016A&A...591A..47L, 2016ApJ...822..109A}, which might be associated with a quasar-driven wind \citep[e.g., ][]{2014MNRAS.445.3878S}.

It may be considered interesting that we do not detect trends of halo bias with color, given that red quasars are more often detected in FIRST than blue quasars, and FIRST-detected quasars at $0.3 < z < 2.3$ have been shown to inhabit more massive dark matter halos ($M_{h}\sim 5\times 10^{13} \ h^{-1} M_{\odot}$) than their radio-quiet counterparts  \citep[$M_{h}\sim 2\times 10^{12} \ h^{-1} M_{\odot}$, ][]{2017A&A...600A..97R}. This may indicate that the halo bias dichotomy in radio-loud/radio-quiet quasars only emerges when examining classical extended radio-loud systems. This idea is consistent with the finding that the differences in radio properties between red and blue systems appear to vanish when considering only systems with extended radio sources \citep{2019MNRAS.488.3109K, 2020MNRAS.494.3061R, 2020MNRAS.494.4802F, 2021MNRAS.505.5283R}. It should be noted however that only $\sim 20\%$ of red quasars are FIRST-detected \citep{2019MNRAS.488.3109K}, which would dilute any signal that may be present.

\begin{figure}
    \centering
    \includegraphics[width=0.45\textwidth]{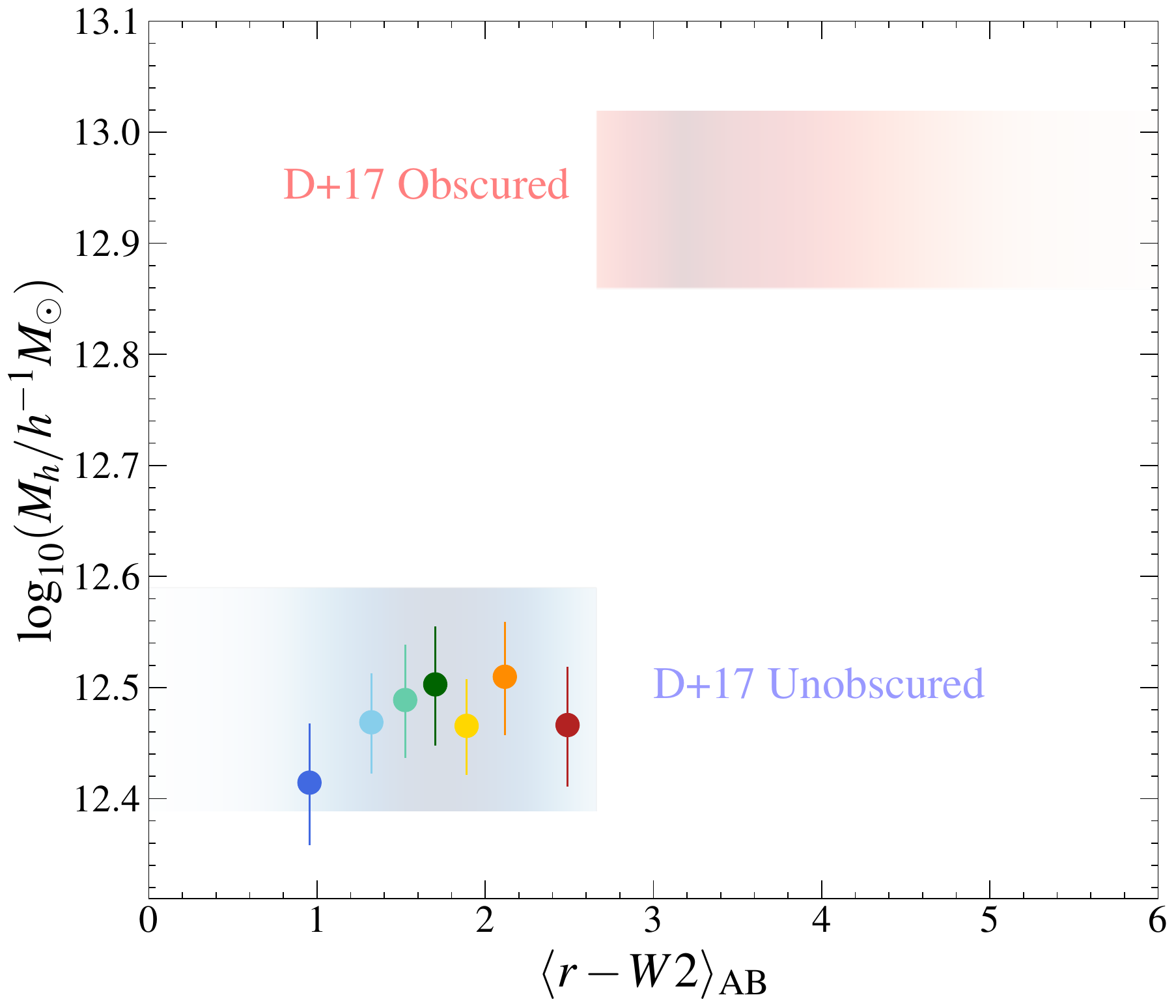}
    \caption{A comparison of the derived characteristic halo masses as a function of median $r-W2$ color of the eBOSS quasar sample (colored circles) and infrared-selected obscured/unobscured AGN from \citet{2017MNRAS.469.4630D} (red/blue gradients, respectively). The vertical span of the gradients represent the $68\%$ confidence interval of the halo masses, while the opacity of the gradient represents the relative number of sources at a given $r-W2$ color in the \citet{2017MNRAS.469.4630D} sample. To derive halo masses for the eBOSS samples, we have binned the eBOSS sample in an identical manner as in Section \ref{sec:binning}, but have used $r-W2$ rather than $g-i$ colors, and then cross-correlated each bin with the entire eBOSS quasar sample. This demonstrates that optically-selected quasars occupy similar halo environments regardless of optical-infrared color, and that their characteristic halo mass is in excellent agreement with infrared-selected unobscured AGN. } 
    \label{fig:masses_w_obscured}
\end{figure}

If quasars' observed optical colors are largely the result of different levels of dust extinction, it may also be considered surprising that we fail to detect trends of clustering with color given that obscured (Type 2) quasars have been shown to inhabit more massive halo environments than their unobscured (Type 1) counterparts \citep{2011ApJ...731..117H, 2014ApJ...789...44D, 2014MNRAS.442.3443D, 2015MNRAS.446.3492D, 2016MNRAS.456..924D, 2017MNRAS.469.4630D, 2018ApJ...858..110P}, which likely indicates a large-scale evolutionary component in the obscuration mechanism \citep{2017MNRAS.464.3526D, 2020ApJ...888...71W}. In order to elucidate this, we repeat our clustering analysis as a function of color using $r-W2$ optical-infrared colors, which have been commonly utilized in the literature as a tracer of AGN obscuration \citep[e.g.,][]{2007ApJ...671.1365H}, rather than $g-i$ colors. We thus perform another cross-correlation analysis of the quasars in each bin with the entire eBOSS quasar sample, and derive bias values and characteristic halo masses exactly as described before. We show the resulting halo masses in Figure \ref{fig:masses_w_obscured}, along with the estimates of the characteristic halo masses of obscured and unobscured infrared-selected AGN from \citet{2017MNRAS.464.3526D}, which utilized both angular clustering and CMB lensing measurements. 

It is clear that eBOSS quasars occupy a small range of host halo mass regardless of optical-mid-infrared color, and that these masses are consistent with the halos surrounding unobscured infrared-selected AGN. It should be noted that the \citet{2017MNRAS.469.4630D} results are from a sample with a wider redshift range $0 < z < 3$, though quasar host halo mass appears to scale weakly with redshift. Given that the colors of infrared-selected quasars appear to be linked to their large-scale environments while optically-selected quasars' colors are not, this may suggest that the location of the obscuring material differs between Type 1 and Type 2 quasars. If obscured AGN and red quasars are simply different stages in a continuous evolutionary sequence, this halo mass difference could alternatively indicate that the obscured phase lasts longer than the transition from obscured to unboscured, with red quasars representing a very brief intermediary stage \citep{2017MNRAS.464.3526D, 2020ApJ...888...71W}. Future work is required to test the halo bias across the full spectrum of color/obscuration to determine the level of extinction at which the halo bias ``turns over'' between Type 1 and 2 quasars, as well as control for redshift, luminosity, and host galaxy properties. This should constrain the nature of the obscuring material surrounding quasars as well as evolutionary models connecting star formation and AGN activity.  

\subsection{Systematics}

Hot, ionized intra-cluster gas can bias CMB lensing measurements by distorting the CMB's spectrum via inverse-Compton scattering, known as the thermal Sunyaev-Zel'dovich \citep[tSZ,][]{1972CoASP...4..173S} effect. We test for a possible tSZ bias to our lensing results by performing a stack in the Planck 2018 SMICA temperature map \citep{2020A&A...641A...4P} at the same positions of quasars. We do not detect any small-scale temperature variation across the color bins expected for tSZ contamination. We do however detect a large-scale ($\sim 30 ^{\circ}$) temperature excess for the blue quasar sample at $\sim 1.5\sigma$. We examine the sky density of our red and blue quasar samples, and find an excess of blue quasars relative to red in the southern galactic hemisphere. This appears to be due to deeper SDSS imaging, which happens to overlap with a large-scale CMB warm spot, generating the large-scale temperature excess observed for blue quasars. This large-scale fluctuation should not affect our estimates of lensing, but we test this by removing all quasars from our sample which lie within HEALPix pixels where the ratio of red to blue quasars differs by more than $1 \sigma$ from the median, and reperform the lensing analysis with this refined sample. Our results are unchanged within the uncertainties, indicating that this temperature fluctuation is not producing dominant systematic errors. Indeed, \citet{2017NatAs...1..795G} showed that lensing estimates are unbiased in the Planck maps even for galaxy clusters with strong tSZ signals.

\section{Conclusions}

We have probed the dark matter halo environments surrounding SDSS quasars as a function of optical color to test whether red and blue quasars occupy different large-scale environments. We have achieved this by utilizing two independent measurements, the two-point cross-correlation functions of eBOSS quasars with three different eBOSS LSS tracers as well as the gravitational lensing of CMB photons around XDQSO quasar candidates. We do not detect any trends of halo bias with color, implying that red and blue quasars trace the underlying matter distribution in a similar manner and occupy dark matter halos of similar characteristic mass, $M_{h}\sim 3\times 10^{12} \ h^{-1} M_{\odot}$. We have also corroborated recent findings demonstrating fundamental differences in the radio properties of red and blue quasars through a stacking analysis of FIRST images, strengthening the conclusion that accretion disk orientation is not the dominant driver of quasar color. These results together appear to suggest that the observed differences arise on nuclear-galactic scales, consistent with recent high-resolution radio-imaging results demonstrating that radio-properties differ between red and blue quasars only when the radio source lies within the host galaxy's extent \citep{2021MNRAS.505.5283R}. We suggest that all of these observations are consistent with a model whereby red quasars' colors are generated by reddening through a nuclear dusty wind launched by the quasar system. Finally, we have shown that optically-selected quasars occupy similar halos across varying $r-W2$ optical-infrared colors, which may help constrain models of AGN and galaxy co-evolution and elucidate the relationship between reddened quasars and heavily obscured AGN.

\acknowledgements

G.C.P. acknowledges support from the Dartmouth Fellowship. R.C.H. acknowledges support from the National Science Foundation through CAREER Award 1554584. D.M.A. and D.J.R. thank the Science, Technology, and Facilities Council (STFC) for funding (grant ST/P000244/1). V.A.F. acknowledges a quota studentship through grant code ST/S505365/1 funded by the STFC. L.K. thanks the Faculty of Science Durham Doctoral Scholarship. Funding for the Sloan Digital Sky Survey IV has been provided by the 
Alfred P. Sloan Foundation, the U.S. 
Department of Energy Office of 
Science, and the Participating 
Institutions. SDSS-IV acknowledges support and 
resources from the Center for High 
Performance Computing  at the 
University of Utah. The SDSS 
website is \url{www.sdss.org}. SDSS-IV is managed by the 
Astrophysical Research Consortium 
for the Participating Institutions 
of the SDSS Collaboration.

\facilities{SDSS, eBOSS, Planck, WISE, JVLA}

\software{astropy: \citet{2018AJ....156..123A}, CAMB: \citet{2000ApJ...538..473L}, colossus: \citet[][]{2018ApJS..239...35D}, Corrfunc: \citet{2020MNRAS.491.3022S}, healpy/HEALPix: \citet[][]{Zonca2019, 2005ApJ...622..759G}}, MANGLE: \citep[][]{2004MNRAS.349..115H, 2008MNRAS.387.1391S}.


\newpage

\bibliography{SDSS_QSO_halos}{}
\bibliographystyle{aasjournal}



\end{document}